\documentclass[10pt]{iopart}

\usepackage{iopams}  
\usepackage{cite} 
\usepackage{hyperref} 
\hypersetup{colorlinks=true, urlcolor=blue, linkcolor=blue, 
citecolor=blue} 
\usepackage{graphics} 
\usepackage{graphicx} 
\usepackage{mathrsfs}
\usepackage{amssymb}

\begin{document}

\title[Electrocaloric effect in the two spin-1/2 XXZ Heisenberg edge-shared tetrahedra]{Electrocaloric effect in the two spin-1/2 XXZ Heisenberg edge-shared tetrahedra and  spin-1/2 XXZ Heisenberg octahedron with Dzyaloshinskii-Moriya interaction}

\author{Hamid Arian Zad}

\address{Alikhanyan National Science Laboratory, Alikhanian Br. 2, 0036 Yerevan, Armenia}%
\address{ICTP, Strada Costiera 11, I-34151 Trieste, Italy}
\eads{\mailto{\normalfont \color{blue} arianzad.hamid@mail.yerphi.am}}

\author{Moones sabeti}%
\address{Young Researchers and Elite Club, Mashhad Branch, Islamic Azad University, Mashhad, Iran}%
\eads{\mailto{\normalfont \color{blue} moones.sabeti@gmail.com}}

\author{Azam Zoshki}%
\address{Young Researchers and Elite Club, Mashhad Branch, Islamic Azad University, Mashhad, Iran}%
\eads{\mailto{\normalfont \color{blue} zoshkia@yahoo.com }}

\author{Nerses Ananikian}%
\address{Alikhanyan National Science Laboratory, Alikhanian Br. 2, 0036 Yerevan, Armenia}%
\address{CANDLE Synchrotron Research Institute, Acharyan 31, 0040 Yerevan, Armenia}
\eads{\mailto{\normalfont \color{blue} ananik@mail.yerphi.am}}

\vspace{10pt}

\begin{abstract}
In the present paper, we consider two species of small spin clusters known as; two spin-1/2 Heisenberg edge-shared tetrahedra and spin-1/2 Heisenberg octahedron with the corresponding Dzyaloshinskii-Moriya terms in a longitudinal magnetic field, then we examine magnetization process and electric polarization of the models as functions of magnetic and electric fields at Low temperature using exact numerical diagonalization. Our exact results are in a good agreement with recent analysis carried out by  J. Stre{\v c}ka and K. Kar{\v l}ov{\' a} \cite{str18b,str17a}.  It is demonstrated that the polarization behavior coincides the magnetization curves (including sequential intermediate plateaus), also it reflects the respective stepwise changes of ground-state phase transitions. We find that the polarization has significant effects on the magnetic field dependencies of the magnetization.  Furthermore, we investigate other isothermal strategies such as cooling rate, magnetocaloric effect (MCE), as well as, electrocaloric effect(ECE) for both models. Since, new electrocaloric materials with high performance is of great interest and importance in condensed matter physics, here, we report on the ECE of the both aforedescribed spin-1/2 Heisenberg small clusters. We conclude that two spin-1/2 Heisenberg edge-shared tetrahedra is a material with significant reversible temperature change capability under an external electric field compared with the spin-1/2 Heisenberg octahedron, and can be used for cooling/heating process. On the other hand, our exact results obtained by studying the magnetocaloric effect and electrocaloric effect of the spin-1/2 Heisenberg octahedron prove that this model displays a rich electrocaloric effect close to the relevant critical points, thus can be considered as a material with rich electrocaloric(magnetocaloric) performance even in the presence of an external electric(magnetic) field. Generally, by comparing both models with each other, we point out that the two spin-1/2 Heisenberg edge-shared tetrahedra is a high performance electrocaloric material rather than spin-1/2 Heisenberg octahedron, while the spin-1/2 Heisenberg octahedron demonstrates greater magnetocaloric effect.
\end{abstract}
\pacs{03.67.Bg, 03.65.Ud, 32.80.Qk \\
{\noindent{\it Keywords}: Magnetization, Specific heat, Magnetocaloric effect, Electrocaloric effect, Cooling rate}}
%
%
%
\section{Introduction}
In condensed matter physics, one of the most investigated subjects is magnetic properties of the single molecular magnets (SMMs) which has attracted  considerable attentions to explore  magnetic behavior of small spin clusters. Magnetic properties of small clusters are strongly linked to their structures, as well as, the character of interaction between their ingredients \cite{Martinez2013,Die2016,str15,kar17,str17a,str18b}. 
The study of effects induced by external electric and magnetic fields on the Heisenberg spin models has been of particular interest, since the magnetic and electric properties of the Heisenberg spin models in the presence of external fields (depending on their direction and magnitude) can be both qualitatively and quantitatively different \cite{Dmitriev12002,Dmitriev2004,Hieida2001,Mori1995,Dmitriev22002,Hagemans2005,Hikihara1998,Hikihara2004,
Kurmann1982,Ovchinnikov2003,Yang11966,Yang21966,Caux2003}.   

Among the most notable features of zero temperature magnetization process, intermediate magnetization plateaus \cite{str18b,str17a,str15,kar17,Honecker2004,Lacroix2011}, quasi-plateaus \cite{Bellucci2014,Ohanyan2015,Ohanyan2018}, magnetization steps and magnetization jumps \cite{str18b,str17a,str15,kar17,Schulenburg2002,Shapira2002} are remarkably considered to study by researchers. Interestingly, magnetic and thermodynamic properties of several insulating magnetic materials can be dramatically described by one-dimensional Heisenberg spin models. For example, Ising-Heisenberg spin models provide a reasonable quantitative description of the thermal and magnetic behaviors associated with the materials with ordered structures in the real world \cite{Antonosyan2009,Paulinelli2013,Lisnyi2016}, diamond chains \cite{Gu2007,Rojas2012,Torrico2014,Ananikian2012,Abgaryan2015,Rojas2017}, magnetic spin ladders \cite{Koga1988,str16,Zad2017,Zad2018} are a class of low-dimensional magnetic materials.

Recently, J. Stre{\v c}ka {\it et al.} have widely reported for the magnetocaloric effects and magnetic properties of various kinds of small clusters known as spin-1/2 XXZ Heisenberg regular polyhedra in references \cite{str17a,str15,kar17}. Then, they have investigated the magnetization process and low-temperature thermodynamics of a special small cluster so-called two spin-1/2 Heisenberg edge-shared tetrahedra in reference \cite{str18b}.
Let us also mention that similar findings have been reported by other researcher, for instance,
  L. M. Volkova and D. V. Marinin have comprehensively worked on a similar set of materials known as corner-sharing $Cu_3$ tetrahedra in three volcanic minerals: averievite $Cu_5O_2(VO_4)_2(CuCl)$, ilinskite $NaCu_5O_2(SeO_3)_2Cl_3$, and avdononite $K_2Cu_5Cl_8(OH)_42H_2O$, and obtained new stimulating results on phase transition, magnetic and electric properties of the aforediscribed materials \cite{Volkova2018,Volkova2017,Volkova12018}. 

The dependence of the isentropy lines in spin models on such Hamiltonian parameters as external fields or exchange interactions gives rise to a variety of caloric effects. Such phenomena are intriguing not only because they provide scientists by deep insights into the physical behavior of various quantum spin clusters by probing their isothermal properties in different situations, but mainly due to high potential for applications in cooling/heating process. The most frequently studied phenomenon in this issue is the MCE, emerging due to dependence of entropy on the magnetic field. The less commonly explored phenomenon is the ECE, which causes by the external electric field effect on the entropy of clusters \cite{Tishin16,Franco16,Correia14,Szalowski18,Littlewood16}. Such effect is recently suggested to possess remarkable advantages when applied in energy-conserving cooling systems. Therefore, a search for novel materials exhibiting giant ECE with optimized parameters is well motivated. In the present work, based on the previous studies \cite{str18b,str17a,str15,kar17}, we will comprehensively examine the magnetization process, polarization curves, MCE and ECE for two kinds of small spin clusters named; two spin-1/2 XXZ Heisenberg edge-shared tetrahedra, and spin-1/2 XXZ Heisenberg octahedron. 

The remainder of this paper is organized as follows.  The quantum spin-1/2 Heisenberg on two edge-shared tetrahedra model and  spin-1/2 Heisenberg octahedron  in the presence of the both magnetic and electric fields are characterized in section \ref{Model}. The most interesting results for the low-temperature magnetization process, polarization, specific heat, cooling rate, MCE and ECE are discussed in section \ref{TM}. Finally, our paper will end up in section \ref{conclusion} with several concluding remarks and future outlooks.
\begin{figure}
\begin{center}
\resizebox{0.45\textwidth}{!}{%
\includegraphics{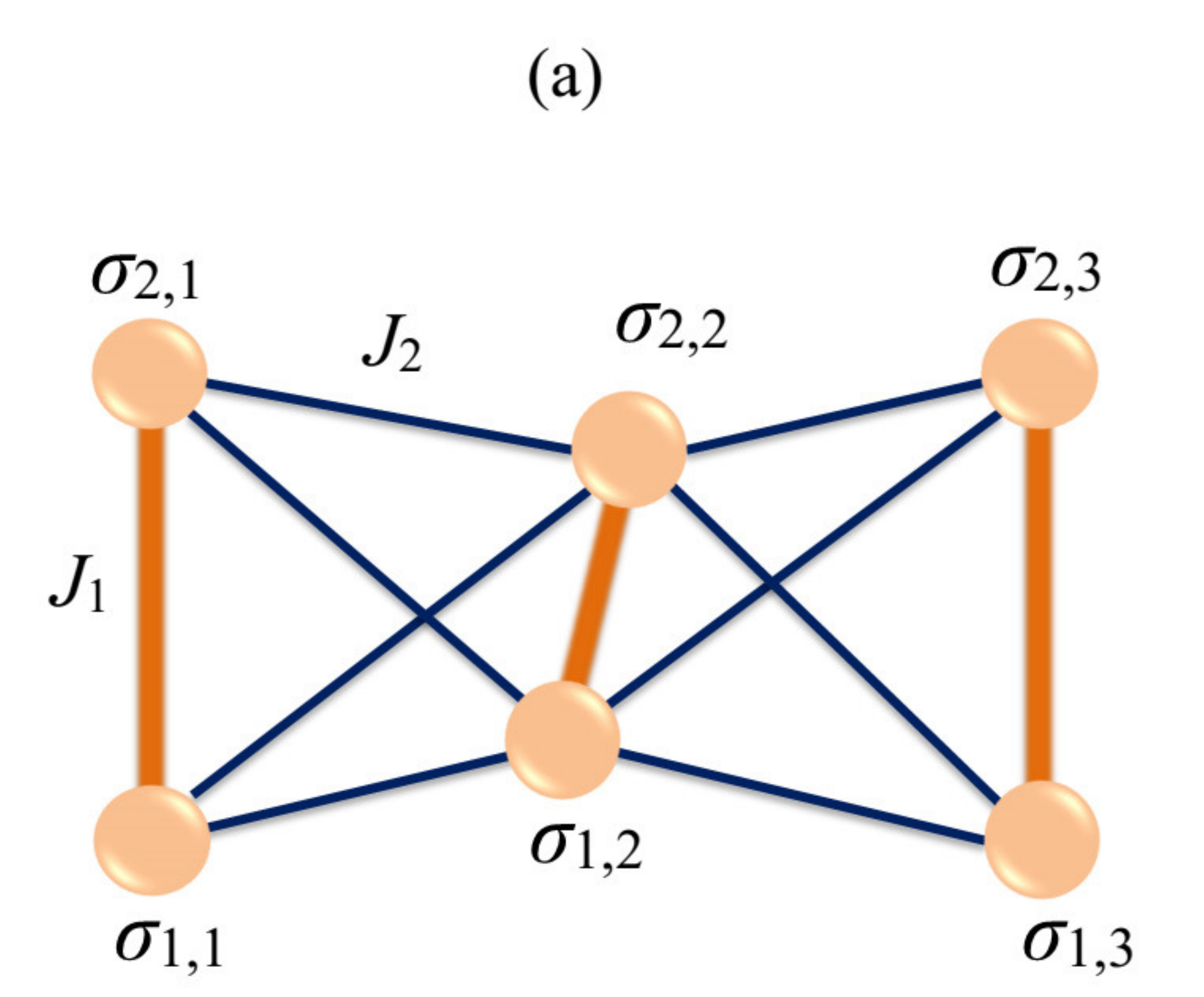}
}
\resizebox{0.45\textwidth}{!}{%
\includegraphics{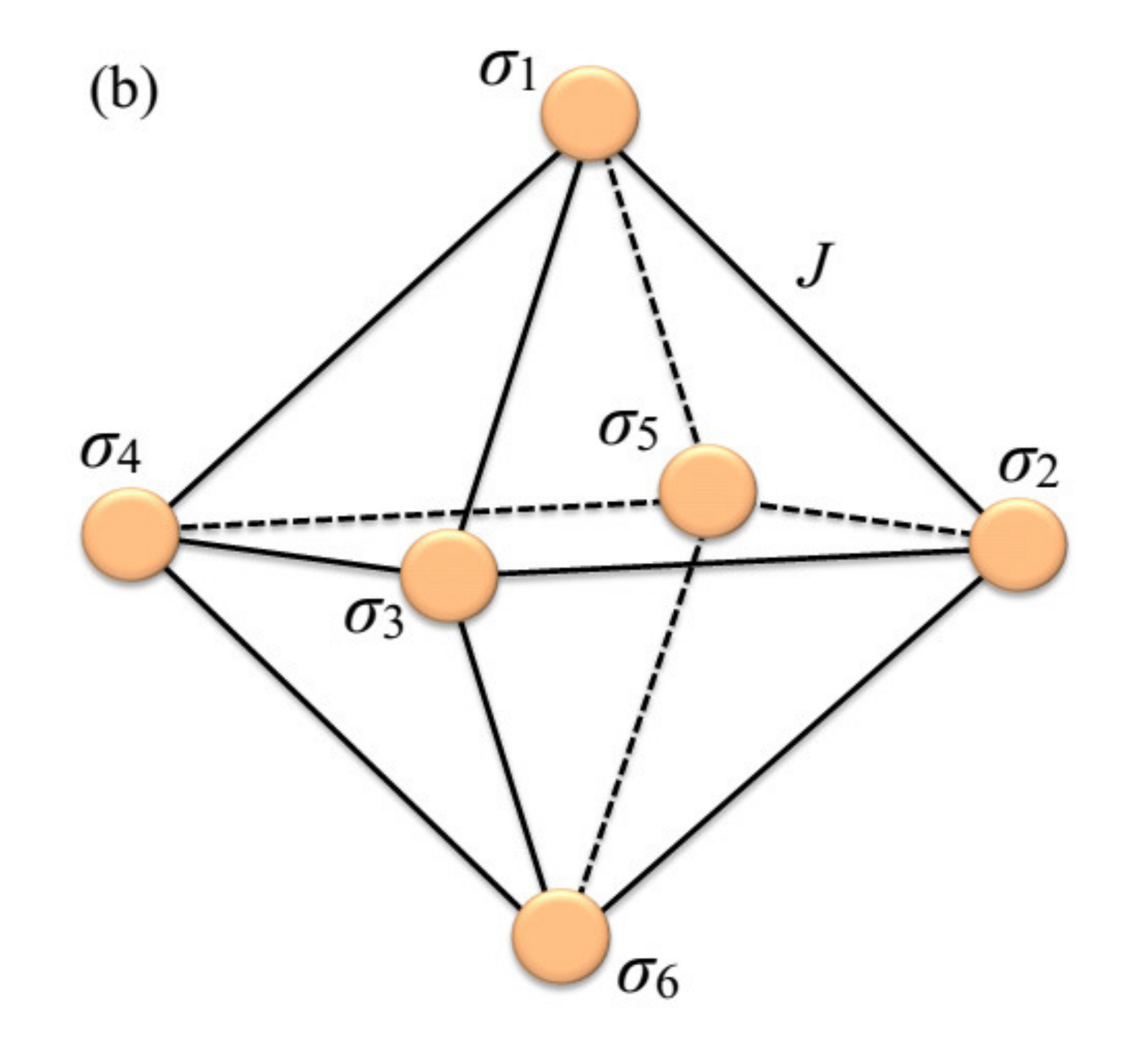}
}
\caption{(a) A schematic representation of the two spin-1/2 edge-shared tetrahedra small cluster.  Thick (orange) lines correspond to the Heisenberg intradimer interactions $J_1$, while thin (Dark-blue) lines represent the Heisenberg interdimer interactions $J_2$. (b) Spin-1/2 Heisenberg octahedron with the nearest-neighbor coupling constant $J$.}

\label{fig:SpinModel_T_O}
\end{center}
\end{figure}
\section{Models and exact numerical solution}\label{Model}
Let us consider the anisotropic spin-1/2 XXZ Heisenberg model on two edge-shared tetrahedra small cluster with the geometry of butterfly-shaped 
(figure \ref{fig:SpinModel_T_O}(a)) in a transverse magnetic field  with a typical Dzyaloshinskii-Moriya (DM) interaction as
\begin{equation}\label{hamiltonian1}
\begin{array}{lcl}
H_1 ={J}_1\sum\limits_{i=1}^3{\boldsymbol S}_{1,i}\cdot{\boldsymbol S}_{2,i}+{J}_2\sum\limits_{i=1}^2({\boldsymbol S}_{1,i}+{\boldsymbol S}_{2,i})\cdot({\boldsymbol S}_{1,i+1}+{\boldsymbol S}_{2,i+1})\\
-g\mu_BB_z\sum\limits_{i=1}^3({S}_{1,i}^z+{S}_{2,i}^z)+
\gamma E_y\sum\limits_{i=1}^3({S}_{1,i}^x{S}_{2,i}^y-{S}_{1,i}^y{S}_{2,i}^x),
\end{array}
\end{equation}
while, the spin-1/2 Heisenberg octahedron can be defined by the following Hamiltonian
\begin{equation}\label{hamiltonian2}
\begin{array}{lcl}
H_ 2={J}\sum\limits_{i=2}^5\big[{\boldsymbol S}_{1}\cdot{\boldsymbol S}_{i}+{\boldsymbol S}_{i}\cdot{\boldsymbol S}_{6}+
{\boldsymbol S}_{i}\cdot{\boldsymbol S}_{i+1}\big]\\
-g\mu_BB_z\sum\limits_{i=1}^6{S}_{i}^z+
\gamma E_y\sum\limits_{i=2}^5\big({S}_{i}^x{S}_{i+1}^y-{S}_{i}^y{S}_{i+1}^x\big).
\end{array}
\end{equation}
 ${S}_{\Theta,i}^\alpha$ denote the standard spin-1/2 Pauli operators, the superscript $\alpha\in\{x,y,z\}$ labels its spatial components, the subscript 
 ${\Theta=1,2}$ characterizes a given leg in Hamiltonian $H_1$. The material-dependent constants $\gamma$ and $g\mu_B$ ($ \mu_B$ is the Bohr magneton) are absorbed into the electric field strength $J_2E = E_y\gamma$ (in equation (\ref{hamiltonian1})) and $JE = E_y\gamma$ (in equation (\ref{hamiltonian2})), and the external uniform magnetic field $B=g\mu_BB_z$, respectively.
$J_1$ labels the Heisenberg intradimer interaction between the pair spins located on each rung, and the coupling constant $J_2$ labels eight
Heisenberg interdimer interactions. First and second summations in equation (\ref{hamiltonian1}) correspond to the anisotropic Heisenberg coupling between the each pair spins interacted together, which is explicitly given by
\begin{equation}
\begin{array}{lcl}
 J_{\Theta}\left({\boldsymbol S}_{\Theta,i}\cdot{\boldsymbol S}_{\Theta,i^{\prime}}\right)_{\Delta_1}=J_{\Theta}\left(S_{\Theta,i}^xS_{\Theta,i^{\prime}}^x+S_{\Theta,i}^yS_{\Theta,i^{\prime}}^y\right)+\Delta_1 S_{\Theta,i}^zS_{\Theta,i^{\prime}}^z,
\end{array}
\end{equation}
while, in Hamiltonian (\ref{hamiltonian2}) the anisotropic Heisenberg exchange interaction can be identified as 
\begin{equation}
\begin{array}{lcl}
 J\left({\boldsymbol S}_{i}\cdot{\boldsymbol S}_{i^{\prime}}\right)_{\Delta_2}=J\left(S_{i}^xS_{i^{\prime}}^x+S_{i}^yS_{i^{\prime}}^y\right)+\Delta_2 S_{i}^zS_{i^{\prime}}^z,
\end{array}
\end{equation}
for which $\Delta_1$ and $\Delta_2$ denote the anisotropy parameters related to, respectively, $H_1$ and $H_2$. Here, for simplicity we limit ourselves to the case $\Delta_1=\Delta_2=\Delta$. The DM interaction is conventionally considered for the intradimer spins.
It is worthwhile to remark that  Hamiltonians (\ref{hamiltonian1}) and (\ref{hamiltonian2})  without DM terms have been exactly diagonalized in a straight form in the frame of total spin operator of the whole spin clusters $S_T = \sum_{i=1}^6{\boldsymbol S}_i$ and the total spin of each non-interacted dimers in Refs. \cite{str18b,str17a}.

 Since, Hamiltonians (\ref{hamiltonian1}) and (\ref{hamiltonian2}) consist of an extra term DM interaction indicating the existence of an electric field, we use exact numerical calculation to obtain all eigenvalues of the Hamiltonians. It is quite understandable that exact numerical procedure is one of the most valid methods to diagonalize Hamiltonians of the small spin clusters that will be the core of our calculations and simulations in this paper.
 
 Let us explain more about our exact numerical diagonalization method that might be useful for readers. In this procedure we consider the Hamiltonian of clusters as a float matrix, such that its coefficients can be altered by numerical steps. Consequently, we can obtain exact eigenvalues of the aforedescribed Hamiltonian in each step. The so-called modulated Hamiltonian $\mathcal{H}_{mod}$ according to the number of steps for each two parameters, as a representative example in the ($B-T$) plane, where $B$ assumed with $N$ steps ($\{0, n_1, 2n_1,\cdots, N\}$) and $T$ assumed with $M$ steps ($\{0, m_1, 2m_1,\cdots, M\}$) is formulated as 
\begin{equation}\label{NumH}
\mathcal{H}_{mod}= \left(
\begin{array}{cccc}
  \Big[H\Big]_{\scriptscriptstyle{B=0,T=0}} &  \Big[H\Big]_{\scriptscriptstyle{B=0,T=m_1}}  & \cdots &  \Big[H\Big]_{\scriptscriptstyle{B=0,T=M}} \\
  \Big[H\Big]_{\scriptscriptstyle{B=n_1,T=0}}  &   \Big[H\Big]_{\scriptscriptstyle{B=n_1,T=m_1}}  & \cdots & 
   \Big[H\Big]_{\scriptscriptstyle{B=n_1,T=M}}  \\
  \vdots  & \cdots & \ddots & \vdots \\
   \Big[H\Big]_{\scriptscriptstyle{B=N,T=0}}  & \Big[H\Big]_{\scriptscriptstyle{B=N,T=m_1}} & \cdots &  \Big[H\Big]_{\scriptscriptstyle{B=N,T=M}}  \\
 \end{array} \right)_{64N\times 64M}.
\end{equation}
 $n$ and $m$ are arbitrary real numbers that determine the length of each step. Each coefficient of the matrix $\mathcal{H}_{mod}$ is a $64\times 64$ square matrix denoting the corresponding Hamiltonian matrix (here $H_1$ or $H_2$) when the temperature and the magnetic field are taken as special digits $T=n$ and $B=m$, respectively. Thus, each $64\times 64$ coefficient has 64 eigenvalues that would make a single partition function at coordinate ($T=n$, $B=m$). 
 Consequently,  in terms of our numerical scenario, $N\times M$  matrix $\mathcal{Z}$ that includes all partition functions deducted from diagonalizing coefficients of the matrix $\mathcal{H}_{mod}$ can be represented as
 \begin{equation}\label{NumZ}
\mathcal{Z}= \left(
\begin{array}{cccc}
 Z_{\scriptscriptstyle{B=0,T=0}} & Z_{\scriptscriptstyle{B=0,T=m_1}}  & \cdots &  Z_{\scriptscriptstyle{B=0,T=M}} \\
  Z_{\scriptscriptstyle{B=n_1,T=0}}  &  Z_{\scriptscriptstyle{B=n_1,T=m_1}}  & \cdots & 
   Z_{\scriptscriptstyle{B=n_1,T=M}}  \\
  \vdots  & \cdots & \ddots & \vdots \\
   Z_{\scriptscriptstyle{B=N,T=0}}  & Z_{\scriptscriptstyle{B=N,T=m_1}} & \cdots & Z_{\scriptscriptstyle{B=N,T=M}}  \\
 \end{array} \right)_{N\times M}.
\end{equation}

 Next, to obtain a geometrical description of magnetization, polarization and the specific heat, we use Euler$^,$s method that is a numerical procedure to solve differential equations with a given initial value. It is the most basic explicit method for numerical integration of ordinary differential equations and is the simplest Runge-Kutta method \cite{Euler}. 
 Using the thermodynamic relations, the magnetization $M$, polarization $P$, entropy $S$ and the specific heat $C$ can be defined as
 \begin{equation}\label{TParameters}
\begin{array}{lcl}
{M}=-\Big(\frac{\partial f}{\partial B}\Big)_{T}, {P}=-\frac{1}{J}\big(\frac{\partial f}{\partial E}\big)_{T}, {C}=-T\Big(\frac{\partial^2 f}{\partial T^2}\Big)_{B,E_y}, 
 S=\Big(\frac{\partial f}{\partial T}\Big)_{B,E_y}.
 \end{array}
\end{equation}
$f=-k_BT\ln Z$ is the Gibbs free energy, where according to the equation (\ref{NumZ}) is also a  $N\times M$ matrix.

The MCE ($\big(\frac{\partial T}{\partial B}\big)_S=-\big(\frac{(\partial S / \partial B)_T}{(\partial S / \partial T)_B}\big)$) and the ECE ($\big(\frac{\partial T}{\partial E_y}\big)_S=-\big(\frac{(\partial S / \partial E_y)_T}{(\partial S / \partial T)_{E_y}}\big)$) are particularly large nearby the quantum critical points (QCPs). Generally, to detect and classify QCPs, the corresponding quantity so-called the Gr{\" u}neisen parameter is implemented. Therefore, the MCE and ECE are directly associated to the generalized Gr{\" u}neisen ratios 
\begin{equation}\label{GruneisenRario}
\begin{array}{lcl}
\gamma_B=\frac{1}{T}\big(\frac{\partial T}{\partial B}\big)_{T} =-\frac{1}{T}\Big(\frac{(\partial S / \partial B)_T}{(\partial S / \partial T)_B}\Big), \\
\lambda_{E_y}=\frac{1}{T}\big(\frac{\partial T}{\partial E_y}\big)_{T} =-\frac{1}{T}\Big(\frac{(\partial S / \partial E_y)_T}{(\partial S / \partial T)_{E_y}}\Big).
 \end{array}
\end{equation}
Using thermodynamic relations (\ref{TParameters}), Gr{\" u}neisen ratios ${\gamma}_{B}$ and ${\lambda}_{E_y}$ can be related to the
 cooling rates ${\Gamma}_{B}$ and ${\Lambda}_{E_y}$ by multiplying the tempetature $T$ as
 \begin{equation}\label{Coolingrates}
\begin{array}{lcl}
\quad {\Gamma}_{B}=T{\gamma}_{B}=-\frac{T}{C_{E_y, B}}\big(\frac{\partial M}{\partial T}\big)_{E_y, B},\\
{\Lambda}_{E_y}=T{\lambda}_{E_y}=-\frac{T}{C_{E_y, B}}\big(\frac{\partial P}{\partial T}\big)_{E_y, B}.
\end{array}
\end{equation}
 Let us deduce the magnetization of the model using numerical method described above.  
\begin{equation}\label{EulMag}
\begin{array}{lcl}
{M}=-\left(\frac{\partial f}{\partial B}\right)_{T} = -\left(\frac{f[i,j+1]-f[i,j]}{B[j+1]-B[j]}\right)_{T},\\
{P}_{H_1}=-\frac{1}{J_2}\left(\frac{\partial f}{\partial E_y}\right)_{T} = -\frac{1}{J_2}\left(\frac{f[i,j+1]-f[i,j]}{E_y[j+1]-E_y[j]}\right)_{T},\\
{P}_{H_2}=-\frac{1}{J}\left(\frac{\partial f}{\partial E_y}\right)_{T} = -\frac{1}{J}\left(\frac{f[i,j+1]-f[i,j]}{E_y[j+1]-E_y[j]}\right)_{T},
\end{array}
\end{equation}
where, for the polarization we assume the considered Hamiltonians in the ($E_y-T$) plane.  And for other thermodynamic parameters one can regularly use equation (\ref{EulMag}). 
\section{Results and discussion}\label{TM}
\subsection{Spin-1/2 XXZ Heisenberg on two edge-shared tetrahedra model}
In this section, we will perform a comprehensive analysis of the most interesting results for the low-temperature magnetic and thermodynamics scenarios, MCE and ECE of the two spin-1/2 edge-shared tetrahedra with the geometry of butterfly-shaped spin localizations by assuming the antiferromagnetic intradimer interaction ($J_2/J_1>0$). 

\begin{figure}
\begin{center}
\resizebox{0.7\textwidth}{!}{%
\includegraphics{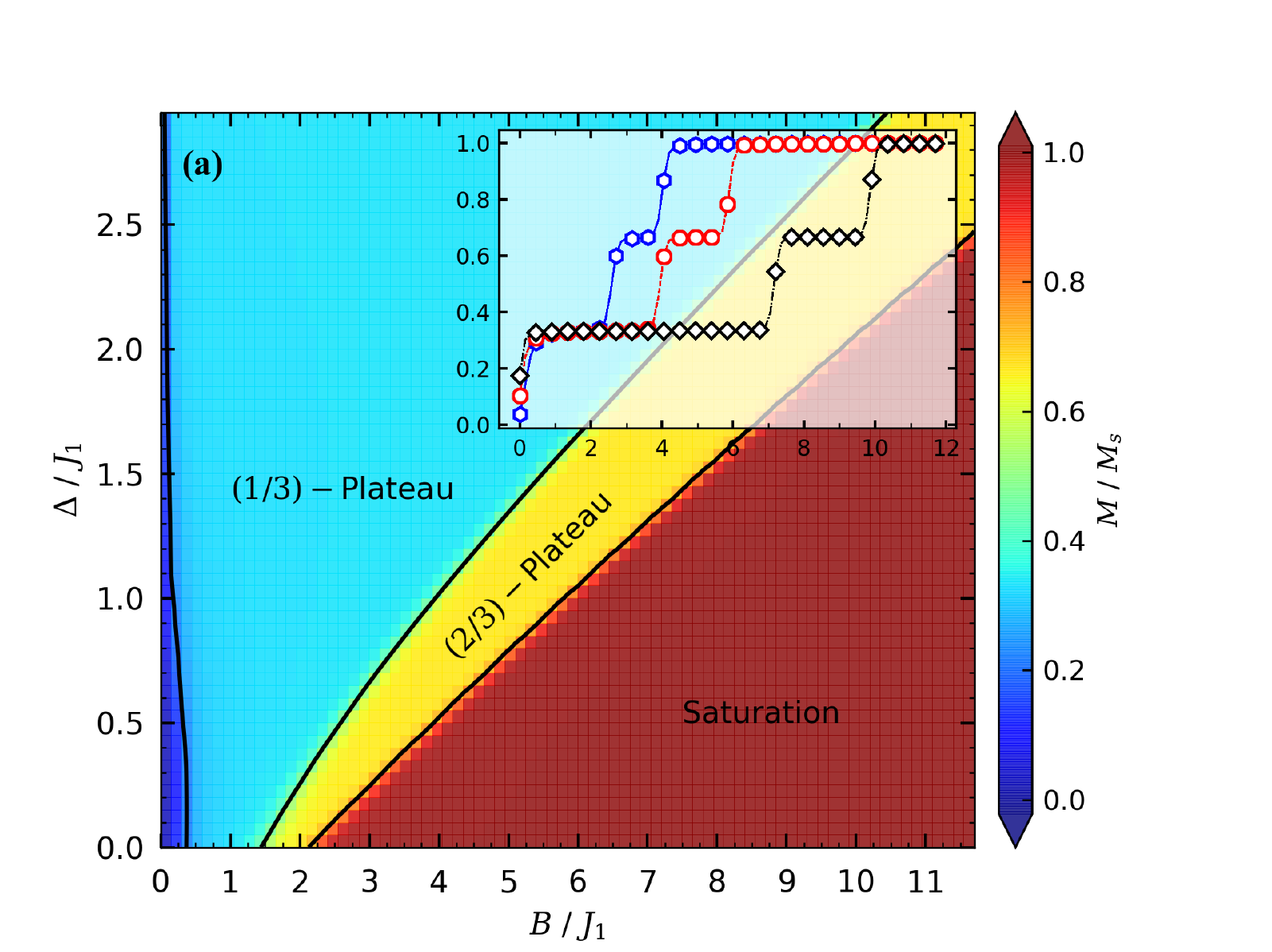}
}
\resizebox{0.7\textwidth}{!}{%
\includegraphics{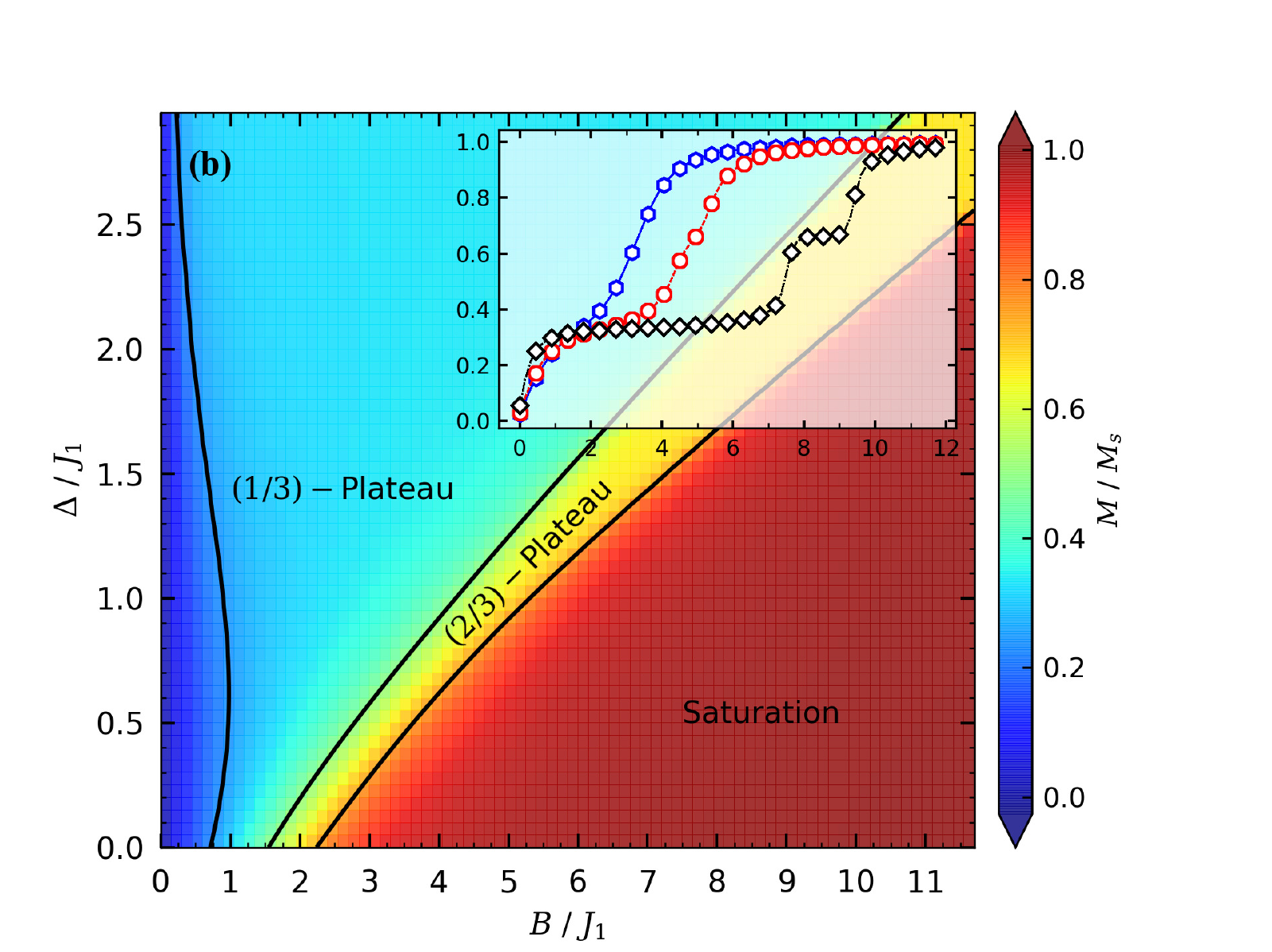}
}
\caption{ Magnetization per saturation value of the tow spin-1/2 edge-shared tetrahedra in the ($\Delta/J_1-B/J_1$) plane at low temperature $T=0.06J_1$ and $J_2=2J_1$ for fixed values of (a) $E_y=0.5J_1$; and (b) $E_y=2J_1$. Insets show the corresponding magnetization curves as a function of the magnetic field for the fixed values of $\Delta=0.5J_1$, $\Delta=J_1$ and $\Delta=2J_1$.}
\label{fig:Mag-BDeltaEy052}
\end{center}
\end{figure}

Heretofore, It was shown that the magnetization curve of the isotropic two spin-1/2 Heisenberg edge-shared tetrahedra versus the magnetic field displays two quite visible intermediate plateaus at $(1/3)$ and $(2/3)$ of the saturation magnetization at low temperature \cite{str18b}. Here, we consider the same spin model but with a Heisenberg anisotropy property $\Delta/J_1$ in the presence of both electric and magnetic fields at low temperature.
Figure \ref{fig:Mag-BDeltaEy052} shows the magnetization of the model under consideration in the ($\Delta/J_1-B/J_1$) plane at low temperature, when we consider a fixed value for the interdimer exchange interaction $J_2$ with respect to the intradimer exchange interaction $J_1$ ($J_2=2J_1$). Panel \ref{fig:Mag-BDeltaEy052}(a) displays the low-temperature magnetization per saturation value $M/M_s$ in the
 ($\Delta/J_1-B/J_1$) plane (as a ground-state phase diagram) when the system is putted in a weak electric field $E_y=0.5J_1$. The intermediate $(1/3)-$plateau and $(2/3)-$plateau,  also saturation magnetization domains are evident in this figure. Black solid lines indicate magnetization jumps from one plateau to one another. With increase of the anisotropy the width of all plateaus monotonically increases as we illustrate in the inset (blue honeycomb-marked line denotes fixed value $\Delta=0.5J_1$, red cycle-marked,  $\Delta=J_1$, and black diamond-marked is for $\Delta=2J_1$).  In fact, boundary between  $(2/3)-$plateau and fully polarized phase can be estimated by the linear equation 
 \begin{equation}\label{DeltaB_Eq1}
\begin{array}{lcl}
\Delta/J_1=\frac{1}{3}B/J_1-\frac{1}{3},
\end{array}
\end{equation}
 and for non-zero anisotropy, boundary between ground state phases associated to the $(1/3)-$plateau and $(2/3)-$plateau is specified by the linear equation
\begin{equation}\label{DeltaB_Eq2}
\begin{array}{lcl}
\Delta/J_1=\frac{1}{4}B/J_1-\frac{1}{2}.
\end{array}
\end{equation}
When the electric field increases (panel \ref{fig:Mag-BDeltaEy052}(b)), magnetization plateaus gradually disappear specifically for the low amounts of anisotropy (red and blue marked lines in the inset of panel \ref{fig:Mag-BDeltaEy052}(b)). On the other hand, we can keep all plateaus alive by tuning the anisotropy, namely by increasing $\Delta/J_1$, magnetization plateaus resist against the electric field ruining effects (black diamond-marked in the inset). Already, all of our achievements have been in a good agreement with the outcomes of the Ref. \cite{kar17}.

\begin{figure}
\begin{center}
\resizebox{0.45\textwidth}{!}{%
\includegraphics{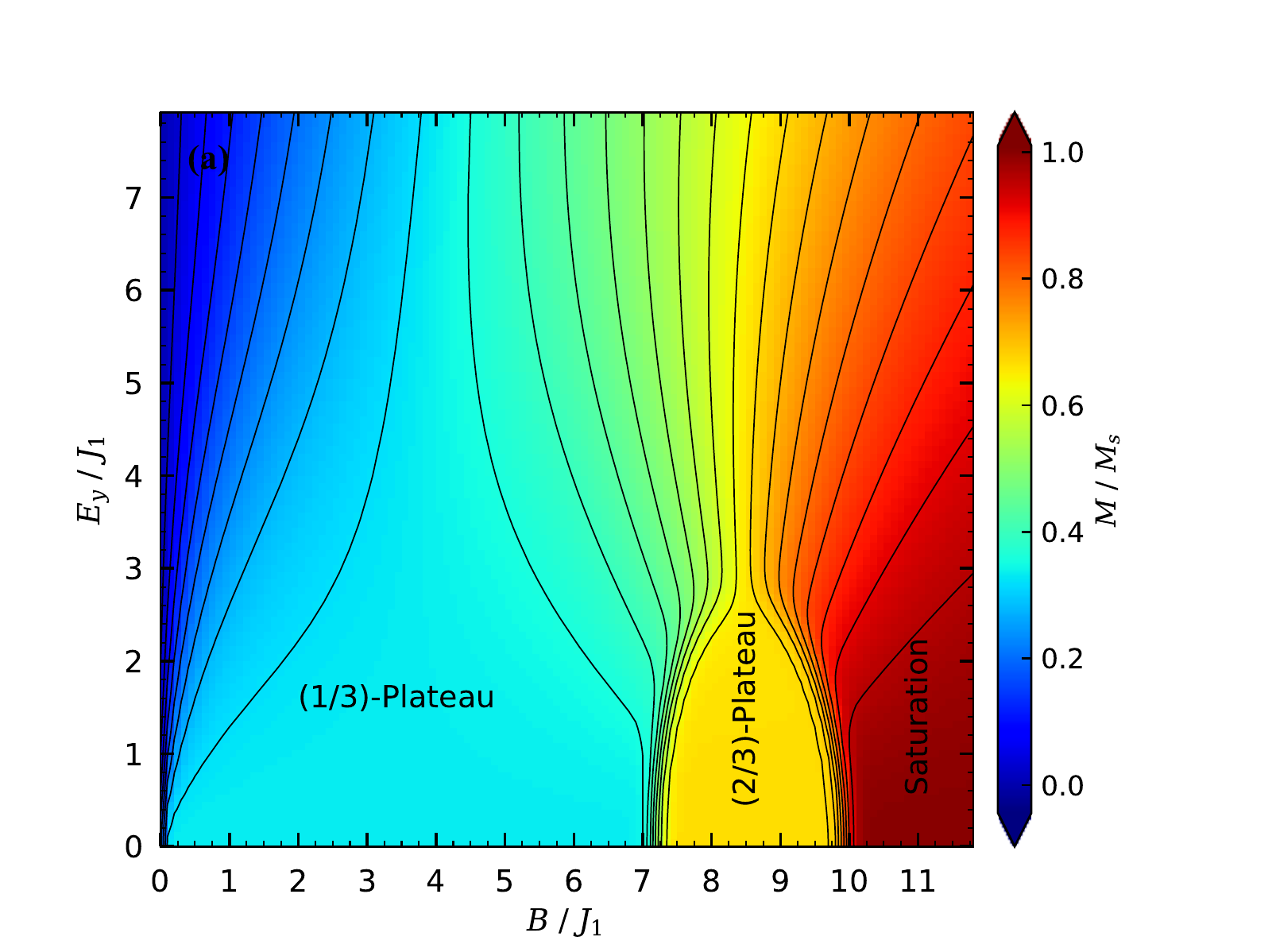}
}
\resizebox{0.45\textwidth}{!}{%
\includegraphics{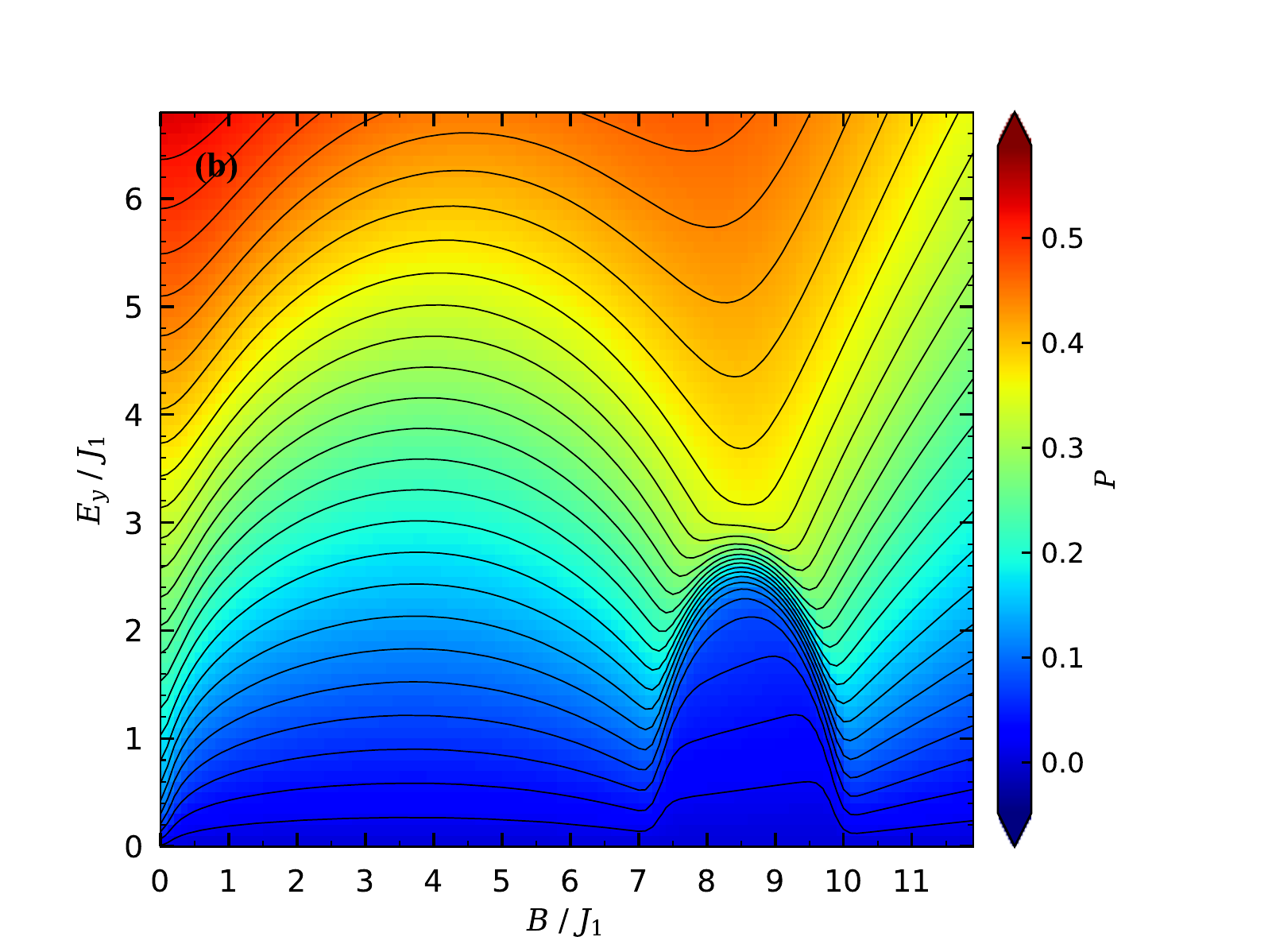}
}
\caption{ (a) Contour plot of the magnetization with respect to its saturation value of the two spin-1/2 XXZ Heisenberg edge-shared tetrahedra model in the ($B/J_1-E_y/J_1$) plane at low temperature $T=0.06J_1$ and fixed $\Delta=J_2=2J_1$. (b) Contour plot of the polarization in the ($B/J_1-E_y/J_1$) plane at low temperature $T=0.06J_1$ and the same fixed values $\Delta=J_2=2J_1$.}

\label{fig:Mag_Pol_BEy}
\end{center}
\end{figure}

ED data for the isothermal magnetization and polarization curves of the two spin-1/2 edge-shared tetrahedra model in the fields 
($B/J_1-E_y/J_1$) plane at low temperature $T=0.06J_1$ and optional fixed values $\Delta=J_2=2J_1$ are confronted in figure \ref{fig:Mag_Pol_BEy}. 
Panel  \ref{fig:Mag_Pol_BEy}(a) depicts the magnetic and electric fields dependencies of the magnetization at finite low temperature. 
The results presented in this figure serve in evidence that the electric field has a substantial effect on the magnetization behavior with respect to the magnetic field. Actually, all plateaus gradually disappear upon increasing the electric field, furthermore, the magnetization will reach its saturation in stronger magnetic fields. 
\begin{figure}
\begin{center}
\resizebox{0.45\textwidth}{!}{%
\includegraphics{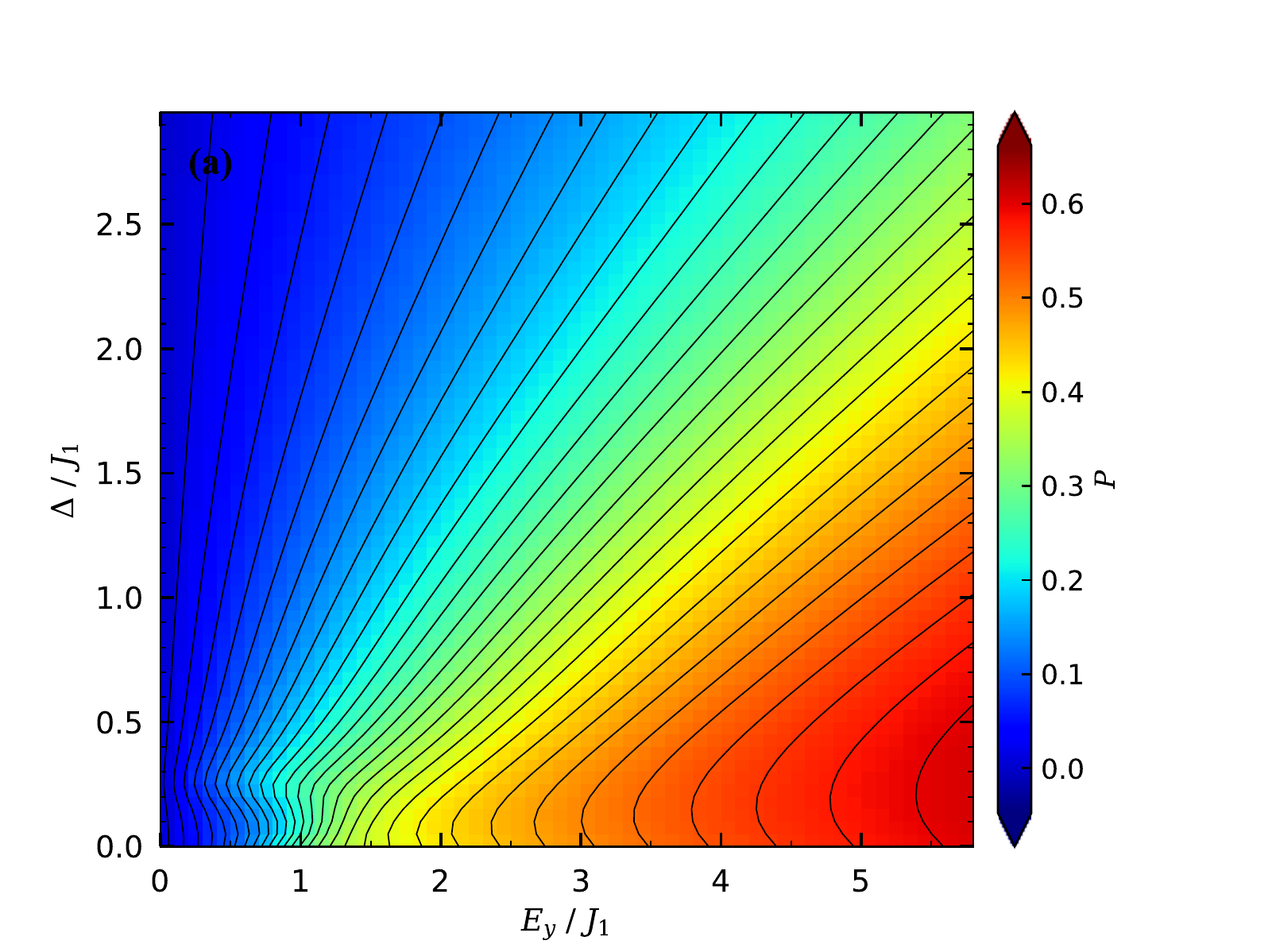}
}
\resizebox{0.45\textwidth}{!}{%
\includegraphics{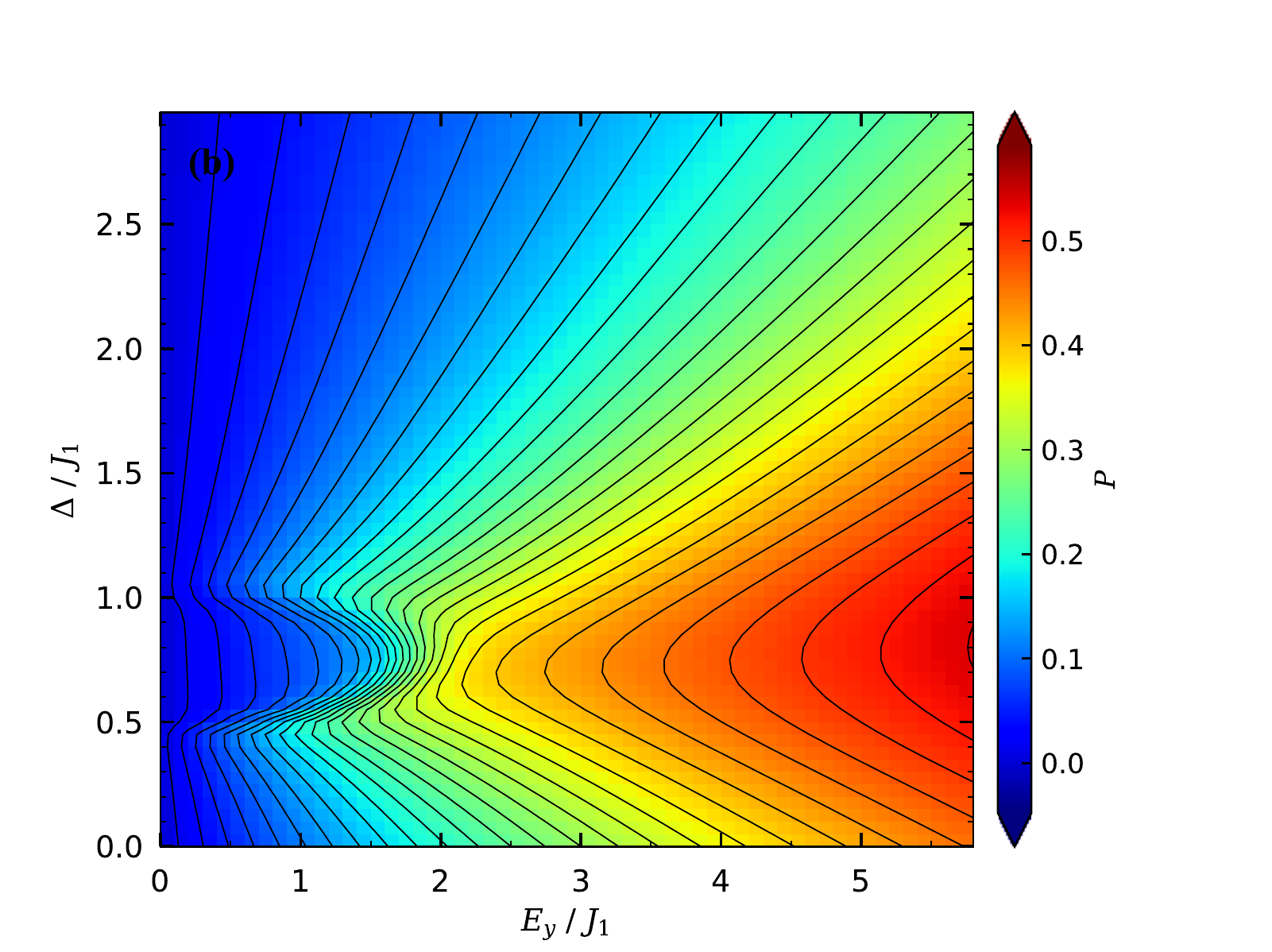}
}
\resizebox{0.45\textwidth}{!}{%
\includegraphics{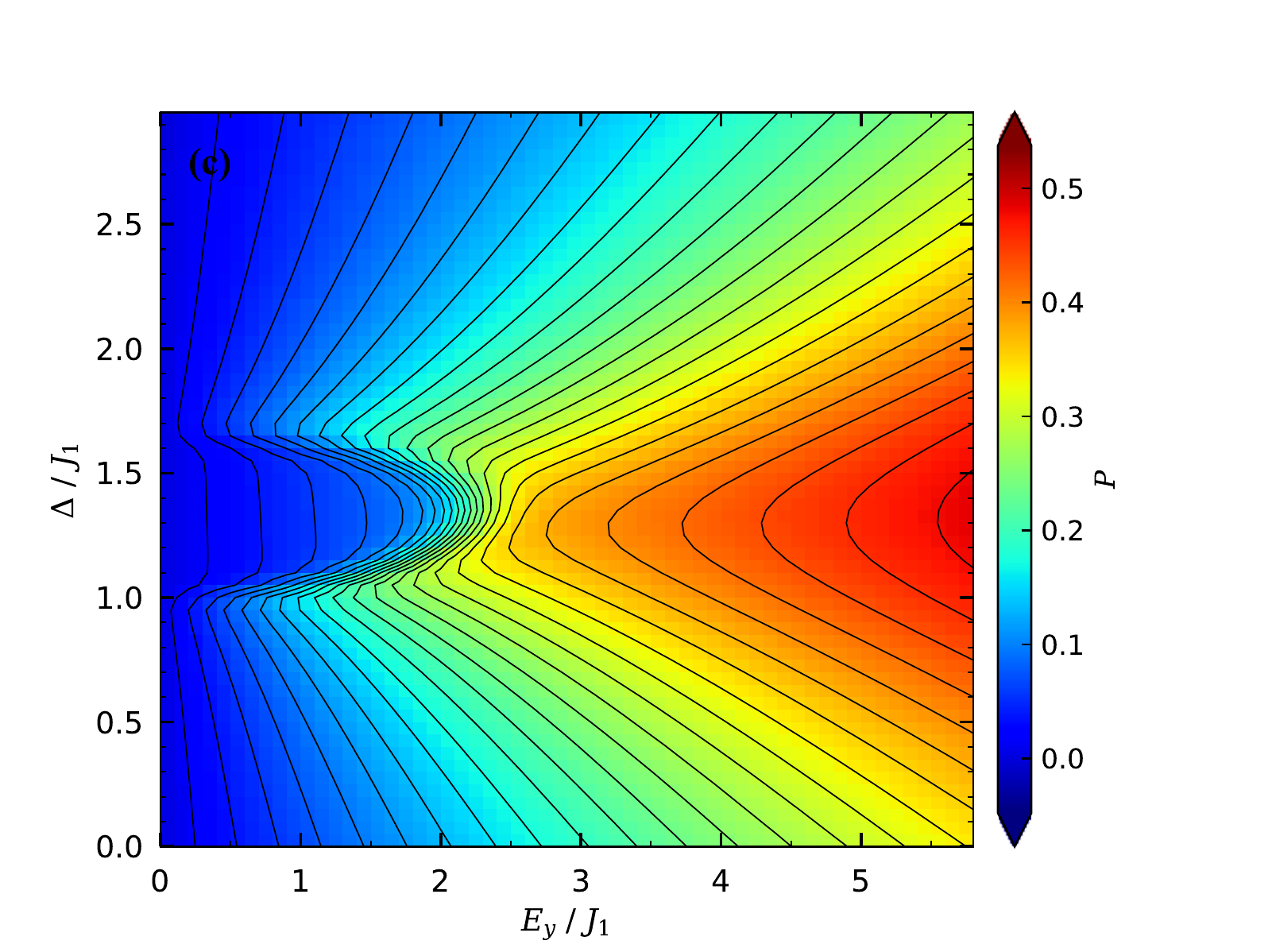}
}
\resizebox{0.45\textwidth}{!}{%
\includegraphics{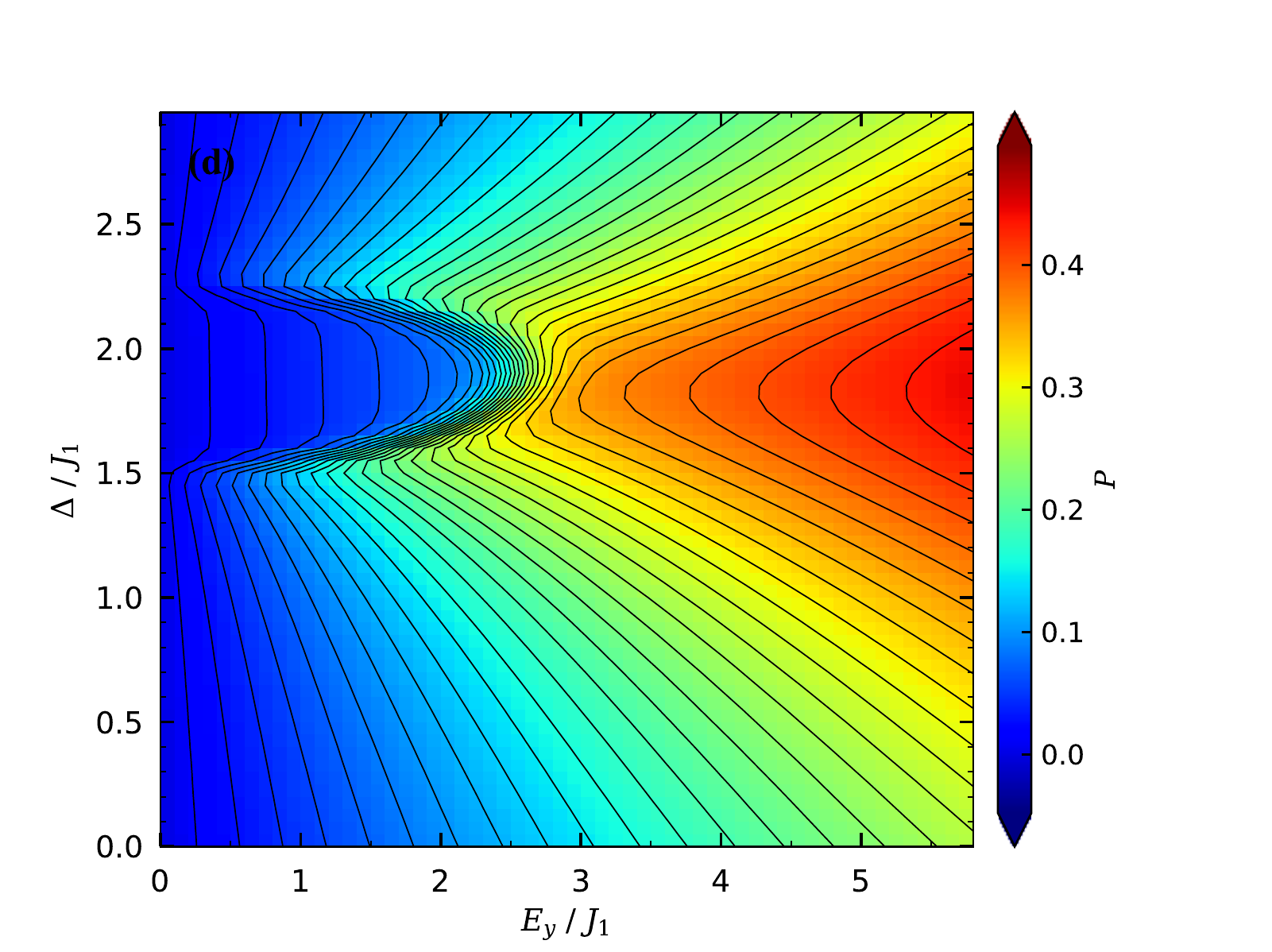}
}
\caption{Contour plots of the polarization of the two edge-shared tetrahedra  in the ($E_y/J_1-\Delta/J_1$) plane at low temperature $T=0.06J_1$ and fixed value $J_2=2J_1$ in the presence of different  magnetic fields: (a) $B=2J_1$; (b)  $B=4J_1$; (c)  $B=6J_1$; and (d)  $B=8J_1$. }

\label{fig:Pol_EyDelta}
\end{center}
\end{figure}

The contour plot of the polarization for the two spin-1/2 XXZ Heisenberg edge-shared tetrahedra model is depicted in figure \ref{fig:Mag_Pol_BEy}(b). 
Interestingly, in accordance with the magnetization plateaus, polarization exhibits an anomalous behavior close to the critical magnetic fields $B=0$, $B=7J_1$ and $B=10J_1$ at which the magnetization jumps occur. With increase of the electric field, the polarization quantitatively increases, as we see a monotonic alteration of the polarization versus rising magnetic field.  Apparently, the low-temperature smoothing of the stepwise magnetization curve remains even for the electric field as strong as $E_y\approx 2.5J_1$, which means intermediate magnetization plateaus persist in the suppressing in the strong electric fields.

The low-temperature effect of the magnetic field increment on the polarization curve of the two spin-1/2 edge-shared tetrahedra model is illustrated in figure \ref{fig:Pol_EyDelta}. It can be understood from this figure that for a particular region of the electric field there are two critical points in the $\Delta/J_1-$axis, at which the polarization non-monotonically changes. By inspecting figure \ref{fig:Mag-BDeltaEy052}, we see that  these critical points denote the magnetization jumps from $(1/3)-$plateau to $(2/3)-$plateau, and in turn  from $(2/3)-$plateau to saturation magnetization. 
In panel  \ref{fig:Pol_EyDelta}(a) is plotted the polarization of the model  in the ($\Delta/J_1-E_y/J_1$) plane for the fixed magnetic field  $B=2J_1$ and $J_2=2J_1$. It is quite clear that the polarization is minimum in the region $\Delta<0.25J_1$ and $E_y<J_1$.  The most remarkable influence of the magnetic field on the polarization behavior is that the mentioned region at which the polarization becomes minimum widely changes upon increasing the magnetic field. For instance, in panel  \ref{fig:Pol_EyDelta}(b) we display the polarization in the ($\Delta/J_1-E_y/J_1$) plane for higher magnetic field $B=4J_1$. Two critical points $\Delta=0.5J_1$ and $\Delta=J_1$ for region $E_y<2J_1$ are visible, revealing the minimum polarization for the model. 
\begin{table}
\begin{center}
\caption{\label{table1} The magnetic field dependence of the critical points in polarization curve.}
\begin{indented}
\item[]\begin{tabular}{@{}llll}
\br
B & $\Delta_1^c$ ($\approx$) & $\Delta_2^c$($\approx$) & $E_y^c$($\approx$) \\
\mr
2$J_1$ & 0 & 0.25$J_1$ & $J_1$\\
4$J_1$ & 0.5$J_1$ & $J_1$ & 2$J_1$ \\
6$J_1$ & $J_1$ & 1.7$J_1$ & 2.5$J_1$\\
8$J_1$ & 1.5$J_1$ & 2.25$J_1$ & 3$J_1$\\
\br
\end{tabular}
\end{indented}
\end{center}
\end{table}
with further increase of the  magnetic field (panels \ref{fig:Pol_EyDelta}(c) and \ref{fig:Pol_EyDelta}(d)), the position of critical anisotropies and  the distance between them monotonically change. Namely, their position shifts toward higher anisotropies,  and meanwhile they move away from each other. Moreover, the critical electric field at which a polarization ramp occurs shifts toward stronger electric fields. This phenomenon indicates that the magnetization $(2/3)-$plateau becomes wider upon increasing the magnetic field. The critical properties of the polarization is shown in table \ref{table1}.

\begin{figure}
\begin{center}
\resizebox{0.45\textwidth}{!}{%
\includegraphics{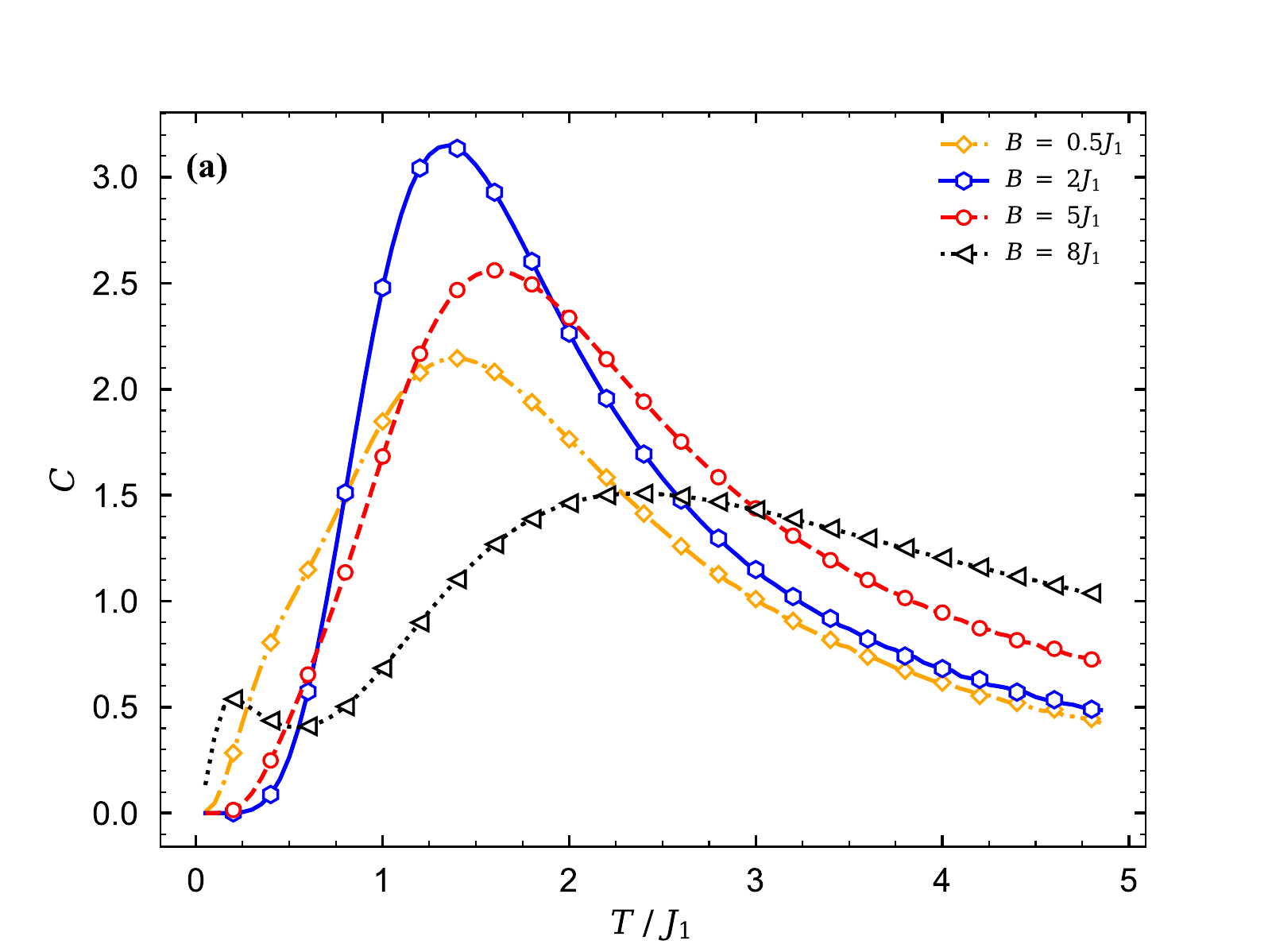}
}
\resizebox{0.45\textwidth}{!}{%
\includegraphics{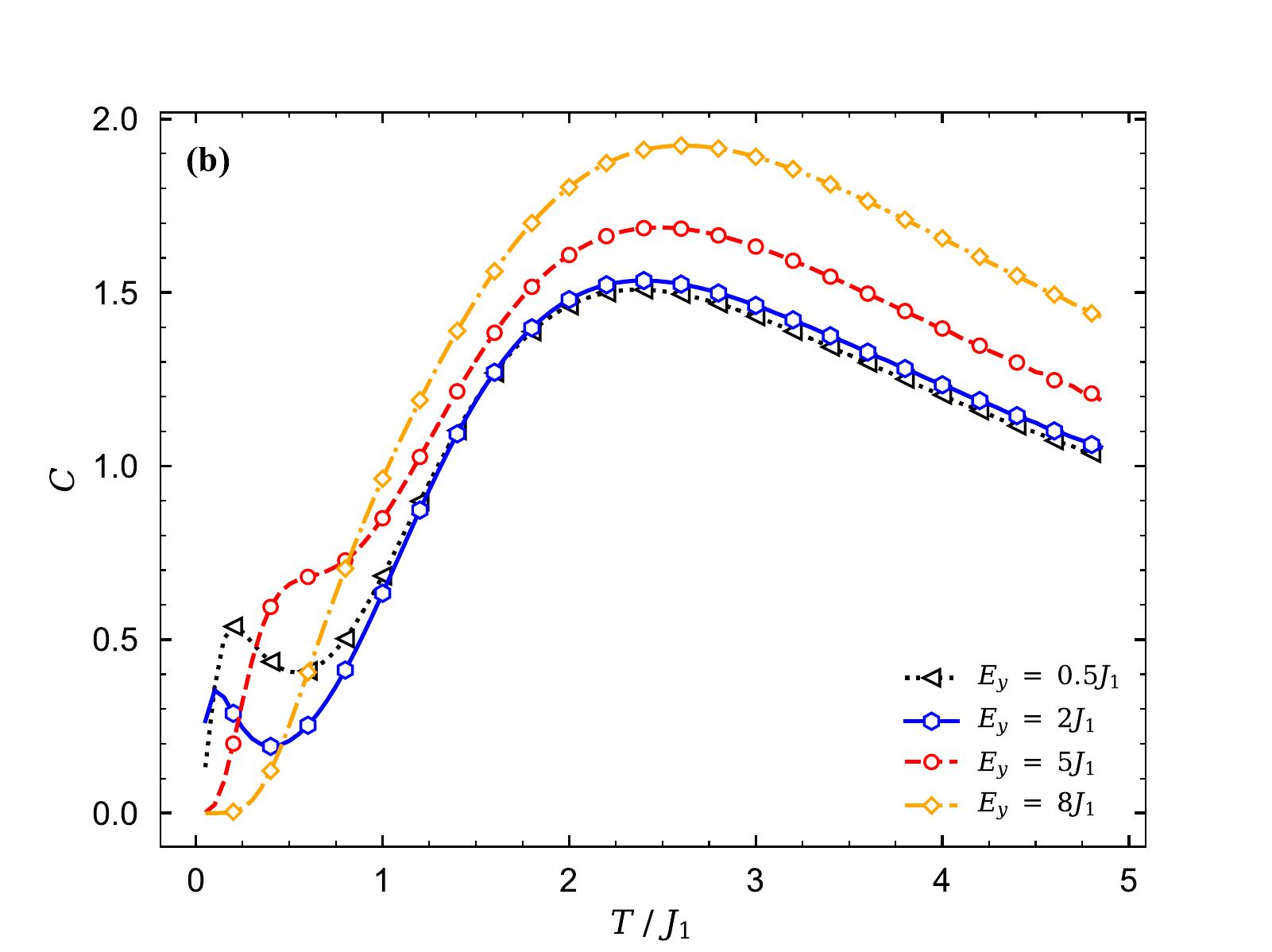}
}
\caption{(a) Specific heat of the two spin-1/2 XXZ Heisenberg edge-shared tetrahedra model as a function of the temperature for several fixed values of the magnetic field, where other parameters have been taken as $E_y=0.5J_1$, $\Delta=J_2=2J_1$. (b) Specific heat as a function of the temperature for several fixed values of the electric field for a set of other parameters as $B=8J_1$, $\Delta=J_2=2J_1$. }

\label{fig:SHeat}
\end{center}
\end{figure}

Last but not least, ED data for the specific heat of the two spin-1/2 Heisenberg edge-shared tetrahedra as a function of temperature for several fixed values of the magnetic field is depicted in figure \ref{fig:SHeat}(a), where other parameters are taken as $E_y=0.5J_1$ and $\Delta=J_2=2J_1$. It is obvious that there is a Schottky-type maximum for the low magnetic fields. The height of Schottky peak alternatively changes with respect to the magnetic field alterations. Nevertheless, with increase of the magnetic field further than $B=5J_1$, the Schottky peak gradually transforms into a double-peak. Indeed, as illustrated in figure \ref{fig:Mag_Pol_BEy} (a), for the strong magnetic fields ($B>5J_1$) and low electric field ($E_y=0.5J_1$), the specific heat curve displays a small anomalous peak due to the zero-temperature phase transition
influence in this rage of electric field.
This scenario is a good evidence of the ground-state phase transition associated to the magnetization jump from $(1/3)-$plateau to $(2/3)-$plateau at critical magnetic field $B_c\approx 7J_1$ in the presence of weak electric fields. 
Although for the higher electric fields ($E>2J_1$) this anomalous peak disappears, because there is no zero-temperature phase
transition in the neighborhood. In panel \ref{fig:SHeat}(b), we show our numerical results for the electric field dependencies of the specific heat as a function of the temperature for strong magnetic fields $B=8J_1$ and $\Delta=J_2=2J_1$.
It is evident from this figure that the specific heat displays a double-peak at sufficiently weak electric fields and strong magnetic fields (black dotted line marked with triangles), which by increasing the electric field merge together and create a Schottky-type peak at higher temperatures. By checking figure  \ref{fig:Pol_EyDelta}(d), we realize that the scenario of the transformation into the Schottky peak upon increasing the electric field is in accordance with the polarization ramp near the electric field $E_y\approx 3J_1$. We quote again that the raising and vanishing second peak in specific heat curve if the result of zero-temperature (magnetic) phase transition in the neighborhood of fixed parameters. 

\begin{figure}
\begin{center}
\resizebox{0.45\textwidth}{!}{%
\includegraphics{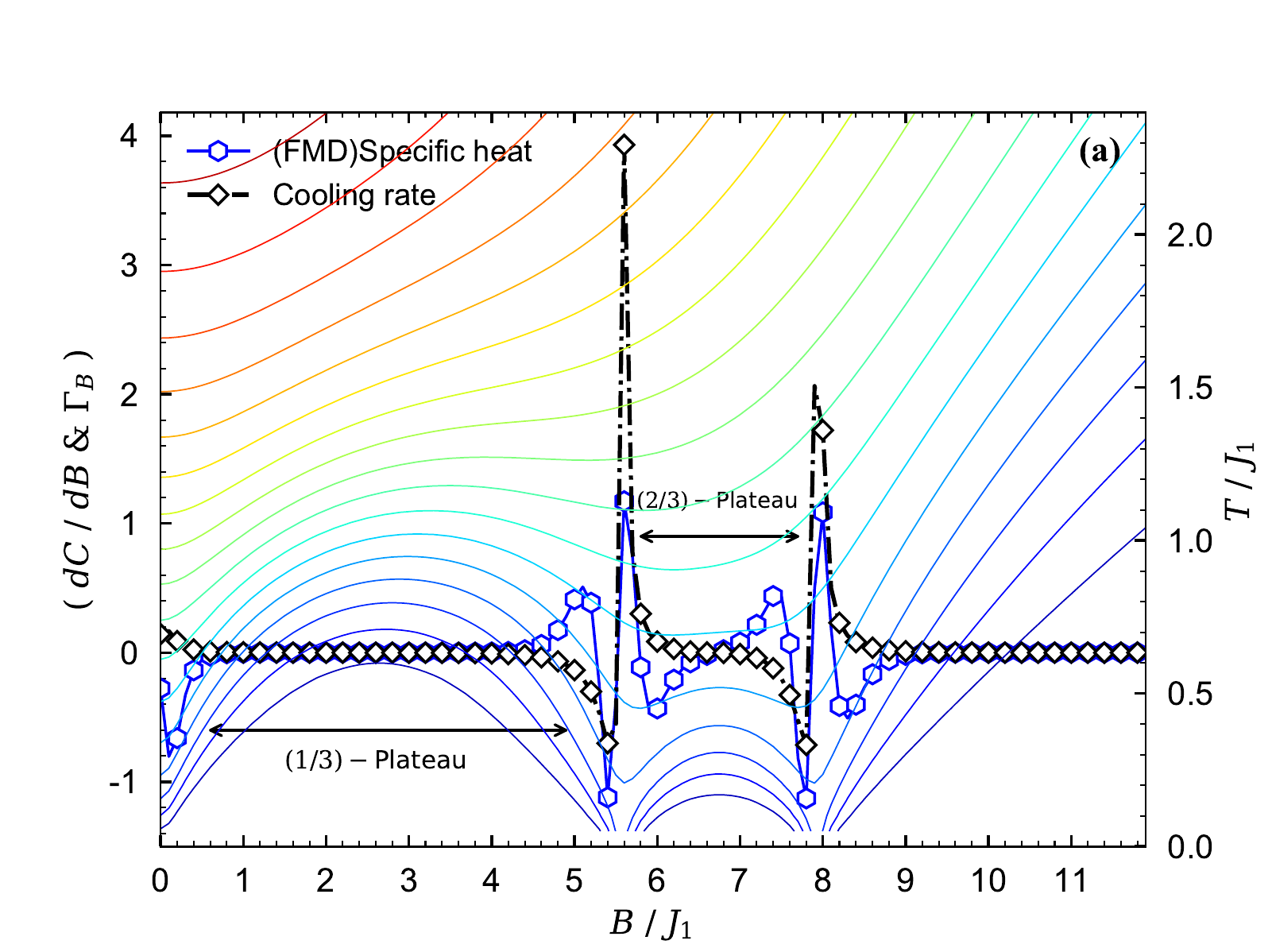}
}
\resizebox{0.45\textwidth}{!}{%
\includegraphics{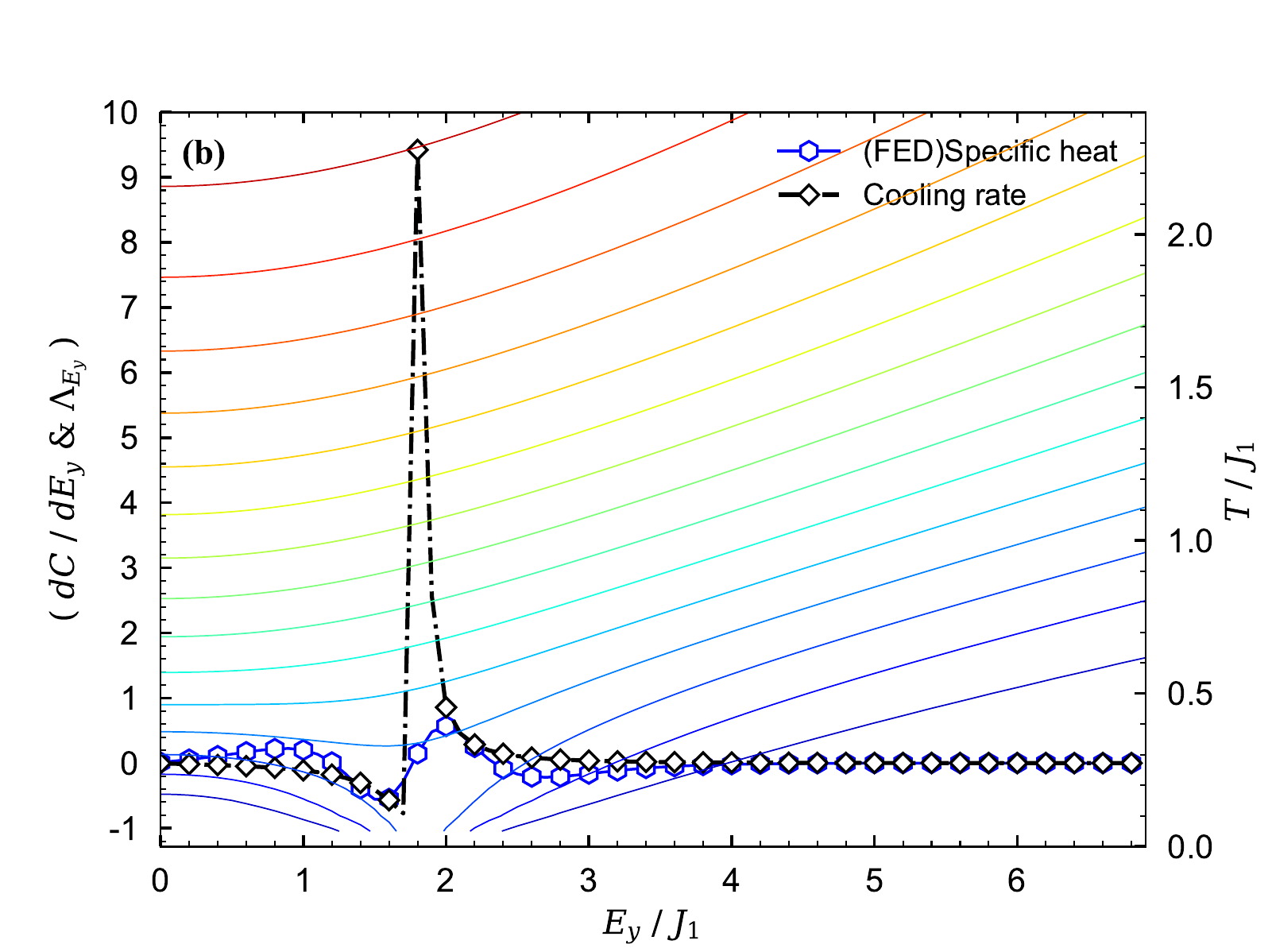}
}
\resizebox{0.45\textwidth}{!}{%
\includegraphics{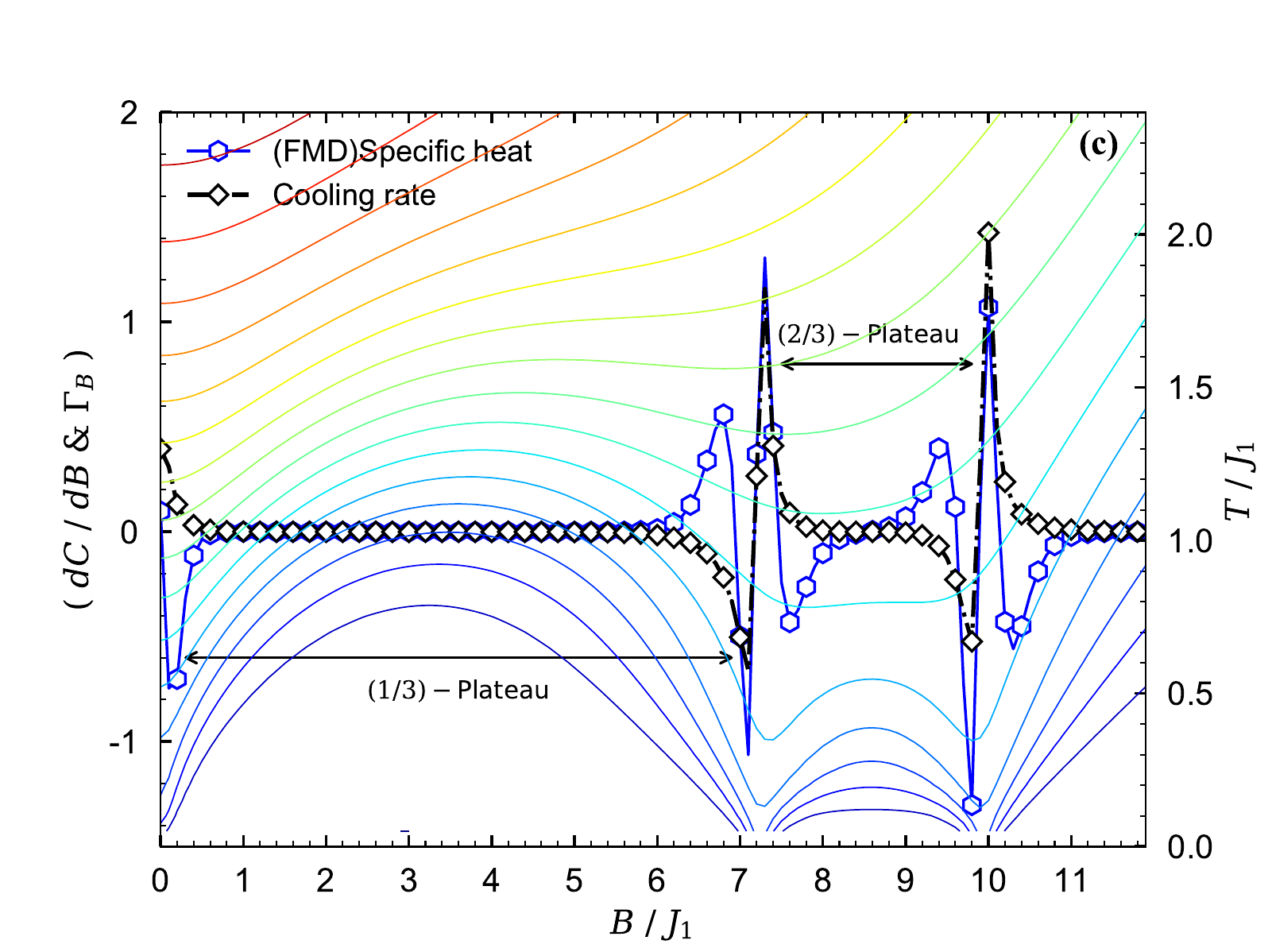}
}
\resizebox{0.45\textwidth}{!}{%
\includegraphics{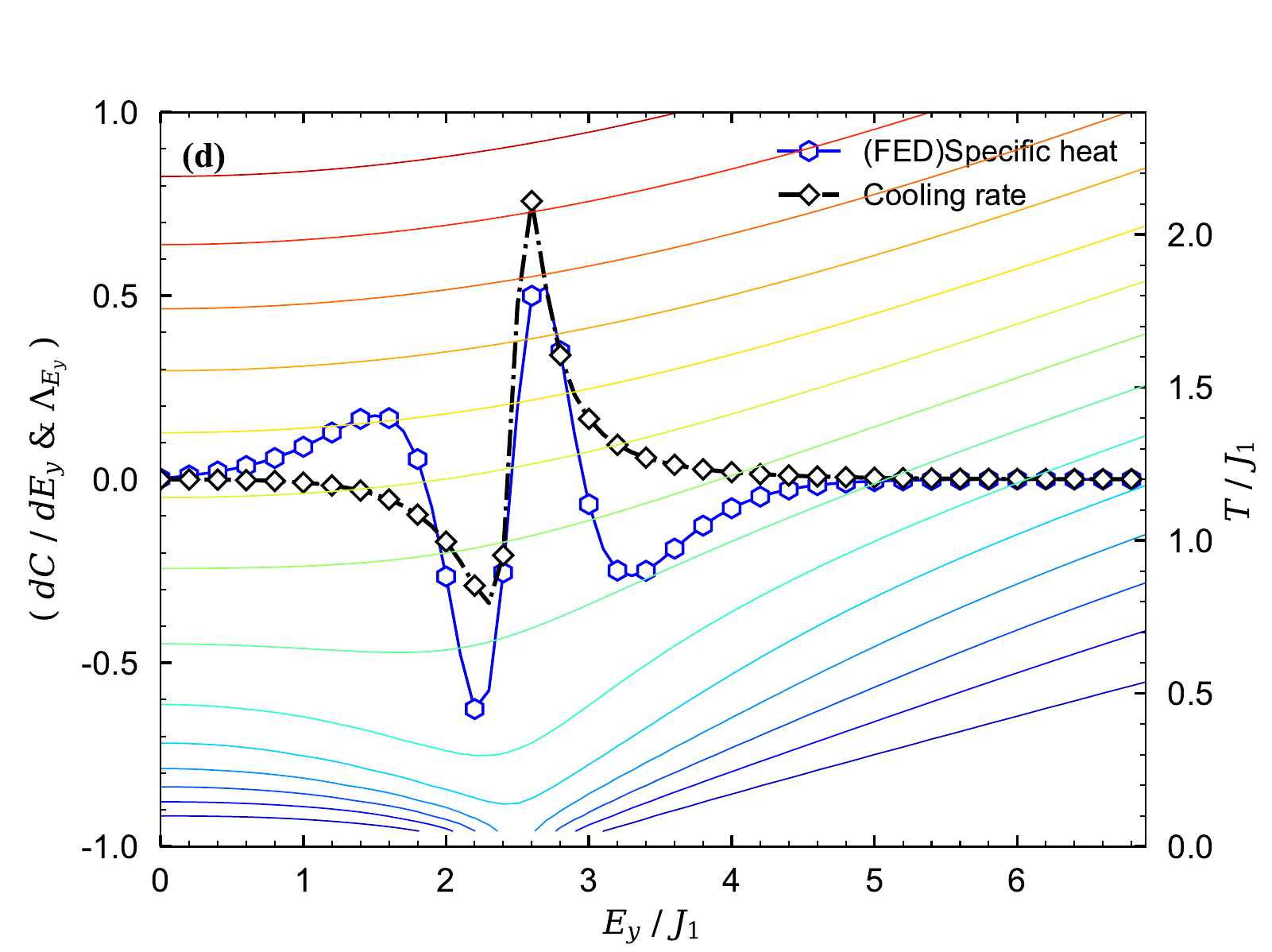} 
}
\caption{The FMD of the specific heat and the cooling rate of the two spin-1/2 Heisenberg  edge-shared tetrahedra model  as functions of the magnetic field at sufficiently low temperature $T=0.06J_1$ and weak electric field $E_y=0.5J_1$ for fixed $J_2=2J_1$, and two different values of the anisotropy: (a) $\Delta=1.5J_1$, (c) $\Delta=2J_1$. Contour plots of the entropy $S$ as a function of the magnetic field and temperature are shown in these panels by assuming the same set of other parameters.
(b) The FED of the specific heat and the cooling rate of the model as functions of the electric field at sufficiently low temperature $T=0.06J_1$ for fixed values of $B=4J_1$, $\Delta=0.75J_2$ and $J_2=2J_1$; (d) $B=8J_1$, $\Delta=2J_2$ and $J_2=2J_1$. Contour plots of the entropy $S$ as a function of the electric field and temperature  are shown in these panels by assuming the same set of other parameters.}
\label{fig:SHeatDiffGamma_BEy}
\end{center}
\end{figure}

In order to accomplish with our discussion concerning to the thermodynamics properties and ground-state phase transition, let us illustrate in figures \ref{fig:SHeatDiffGamma_BEy}(a) and \ref{fig:SHeatDiffGamma_BEy}(c) the first magnetic field derivative of the specific heat and cooling rate 
$\Gamma_B$ of the two edge-shared tetrahedra model as a function of the magnetic field for the low-temperature limit $T=0.06J_1$. The sign change of the cooling rate $\Gamma_B$ close to the critical magnetic field in panels  \ref{fig:SHeatDiffGamma_BEy}(a) and \ref{fig:SHeatDiffGamma_BEy}(c)  clearly indicates a rapid accumulation of the entropy due to the magnetic phase transition between neighboring ground-states $(1/3)-$plateau to $(2/3)-$plateau, and fully polarized state (see black dot-dashed lines marked with triangles plotted for $\Delta=1.5J_1$ in panel \ref{fig:SHeatDiffGamma_BEy}(a) and for $\Delta=2J_1$ in panel \ref{fig:SHeatDiffGamma_BEy}(c), and compare them with the corresponding isothermal dependence of the entropy). By comparing panels  \ref{fig:SHeatDiffGamma_BEy}(a) and \ref{fig:SHeatDiffGamma_BEy}(c) together one can figure out the effect of the anisotropy $\Delta/J_1$ on the enhancement of the MCE in the investigated model. 
It is evident from these figures that  by increasing the anisotropy $\Delta/J_1$, the adiabatic cooling rate $\Gamma_B$ increases close to  the zero field, while it remarkably decreases nearby the other two critical magnetic fields ($B_c=7J_1$ and $B_c=10J_1$ in  panel \ref{fig:SHeatDiffGamma_BEy}(a), and $B_c=5.5J_1$ and $B_c=6J_1$ in panel \ref{fig:SHeatDiffGamma_BEy}(c)). By altering the anisotropy parameter according to the equations (\ref{DeltaB_Eq1}) and  (\ref{DeltaB_Eq2}), the position of the critical magnetic fields will change. Furthermore, one can find out that the cooling rate behavior is in an excellent coincidence with the first magnetic field derivative of the specific heat curves. In other words, figures \ref{fig:SHeatDiffGamma_BEy} (a-d) show, simultaneously, the typical evolution of cooling rates and the first magnetic(electric) field derivative of the specific heat for different applied fields over fixed values of the anisotropy.

Moreover, we display MCE and ECE of the model in figure \ref{fig:SHeatDiffGamma_BEy}, which can be particularly interesting in a vicinity of the magnetization steps and/or polarization ramps. 
Our exact results show a residual entropy nearby the critical magnetic(electric) fields indicating the phase transitions between different magnetic structures gives a rise to an enhanced MCE(ECE) accompanied with a relatively fast cooling of the model during the adiabatic demagnetization (depolarization) process. 

To discuss the MCE in the considered spin-1/2 two edge-shared tetrahedra model, we investigate an adiabatic change of temperature of the model under the magnetic field variation. To this end, a few isentropy lines are plotted in figures \ref{fig:SHeatDiffGamma_BEy}(a) and \ref{fig:SHeatDiffGamma_BEy}(c) in the ($B/J_1-T/J_1$) plane (right vertical axes are the temperature) for the limiting case with $E_y=0.5J_1$ and $J_2=2J_1$,  by assuming  different fixed values of the anisotropy $\Delta=1.5J_1$ and $\Delta=2J_1$. Red region in the isentropy lines corresponds to higher entropy ($S\approx 0.6$), while the blue region corresponds to zero entropy ($S=0$). 
The most surprising finding from panels  \ref{fig:SHeatDiffGamma_BEy}(a) is that the considered model exhibits three enhanced magnetocaloric effect close to the critical fields $B_c=0$,  $B_c=5.5J_1$ and $B_c=8J_1$ at which the abrupt magnetization jump occurs from zero to $(1/3)-$plateau, $(1/3)-$plateau to $(2/3)-$plateau and analogously from $(2/3)-$plateau to saturation magnetization, respectively. As well as, in panel  \ref{fig:SHeatDiffGamma_BEy}(c) three enhanced magnetocaloric effect close to the critical fields $B_c=0$,  $B_c=7J_1$ and $B_c=10J_1$  are observed. As we mentioned before, the change in the position of the critical points at which enhanced magnetocaloric effects are detected, is the result of varying the anisotropy parameter $\Delta/J_1$.

To get a deep insight into the ECE of the Heisenberg two edge-shared tetrahedra model, we have analyzed electrical behavior of the first electric field derivative (FED)  of the specific heat $\frac{d C}{d E_y}$ and the cooling rate $\Lambda_{E_y}$ at low temperature $T=0.06J_1$ and fixed $J_2=2J_1$. The results are separately presented in panel \ref{fig:SHeatDiffGamma_BEy}(b) for the parameter set $B=4J_1$ and $\Delta=0.75J_1$, and in panel \ref{fig:SHeatDiffGamma_BEy}(d)  for the parameter set $B=8J_1$ and $\Delta=2J_1$. A few typical isentropy lines of the model in the ($E_y/J_1-T/J_1$) plane are illustrated as well. It can be seen from  panel \ref{fig:SHeatDiffGamma_BEy}(b) that, the first electric field derivative of the specific heat  $\frac{d C}{d E_y}$ has a peculiar behavior nearby the critical electric field $E_y\approx 1.8J_1$, at which the polarization depicts a huge ramp (review figure \ref{fig:Pol_EyDelta}(d)). Besides, the low-temperature $\Lambda_{E_y}$  curve depicted in this figure (black line marked with diamond) generally increases nearby the critical point $E_y\approx 1.8J_1$ until displays a sharp cusp. According to the figure \ref{fig:Pol_EyDelta}, when we alter the anisotropy $\Delta/J_1$ and magnetic field $B/J_1$ (such that the polarization depicts a ramp versus the electric field), the behavior of both functions $\frac{d C}{d E_y}$ and $\Lambda_{E_y}$  versus the electric field considerably changes. Evidently, the position of the critical electric field shifts toward higher values.

In general,  $\Gamma_B$ ($\Lambda_{E_y}$) and $\frac{d C}{d B(E_y)}$ are similarly positive(negative) close to the intermediate critical pints for cooling(heating) cycle. the characteristic sign change (positive-negative) of cooling rates close to the quantum critical points is due to the accumulation of entropy at the special critical point.

We also display a few isentropy lines in panel \ref{fig:SHeatDiffGamma_BEy}(b) for fixed values $B=4J_1$, $\Delta=0.75J_1$ and $J_2=2J_1$. One may observe an enhanced ECE related to a vigorous change of the external electric field close to a continuous phase transition of the state $(2/3)-$plateau. The displayed isentropy lines can be viewed as the adiabatic temperature response of ECE achieved upon variation of the anisotropy parameter and the magnetic field.
Figure \ref{fig:SHeatDiffGamma_BEy}(d) shows isentropy lines in the ($E_y/J_1-T/J_1$) plan by assuming different fixed values $B=8J_1$, $\Delta=2J_1$ and $J_2=2J_1$. The anisotropy parameter $\Delta/J_1$ and the magnetic field changes lead to increasing the value of the critical electric field at which enhanced ECE is observed.
Although, it turns out that altering the anisotropy part of XXZ exchange interaction suppresses the spectacular magnetocaloric and electocaloric feature, one can compensate this suppressing by intellectual varying of magnetic and electric fields. Namely, electrocaloric response of the
two spin-1/2 XXZ Heisenberg edge-shared tetrahedra model basically depends on a suitable choice of the exchange anisotropy $\Delta/J_1$ and magnetic field $B/J_1$ with the help of figures \ref{fig:Mag_Pol_BEy} and \ref{fig:Pol_EyDelta}.

Consequently, using the adiabatic demagnetization(depolarization) of the two edge-shared tetrahedra model might be of substantial technological importance to refrigeration process, since one can satisfactorily reach ultra-low temperatures in a rather wide range of the magnetic(electric) fields.

\subsection{Spin-1/2 XXZ Heisenberg octahedron}
\begin{figure}
\begin{center}
\resizebox{0.45\textwidth}{!}{%
\includegraphics{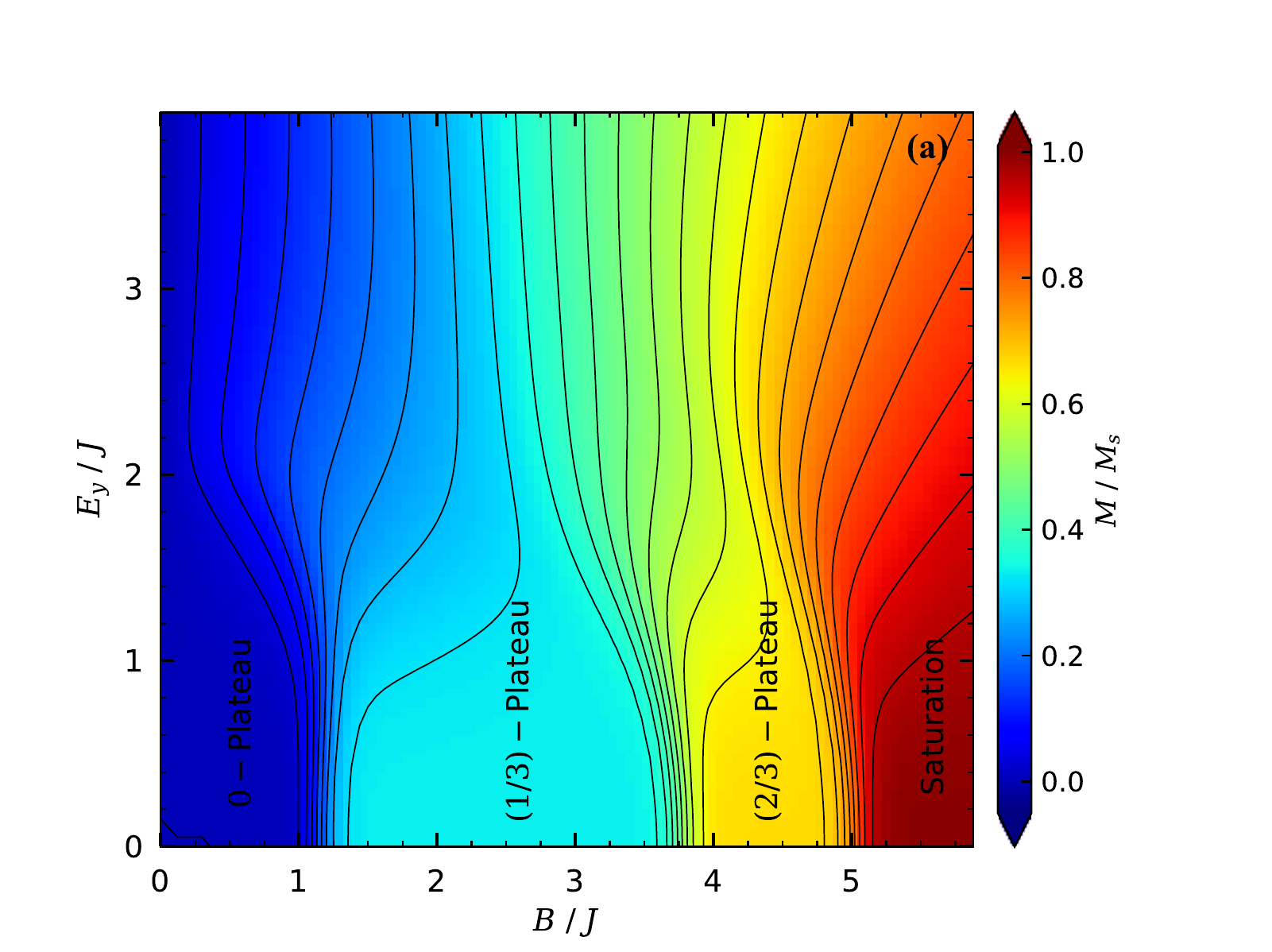}
}
\resizebox{0.45\textwidth}{!}{%
\includegraphics{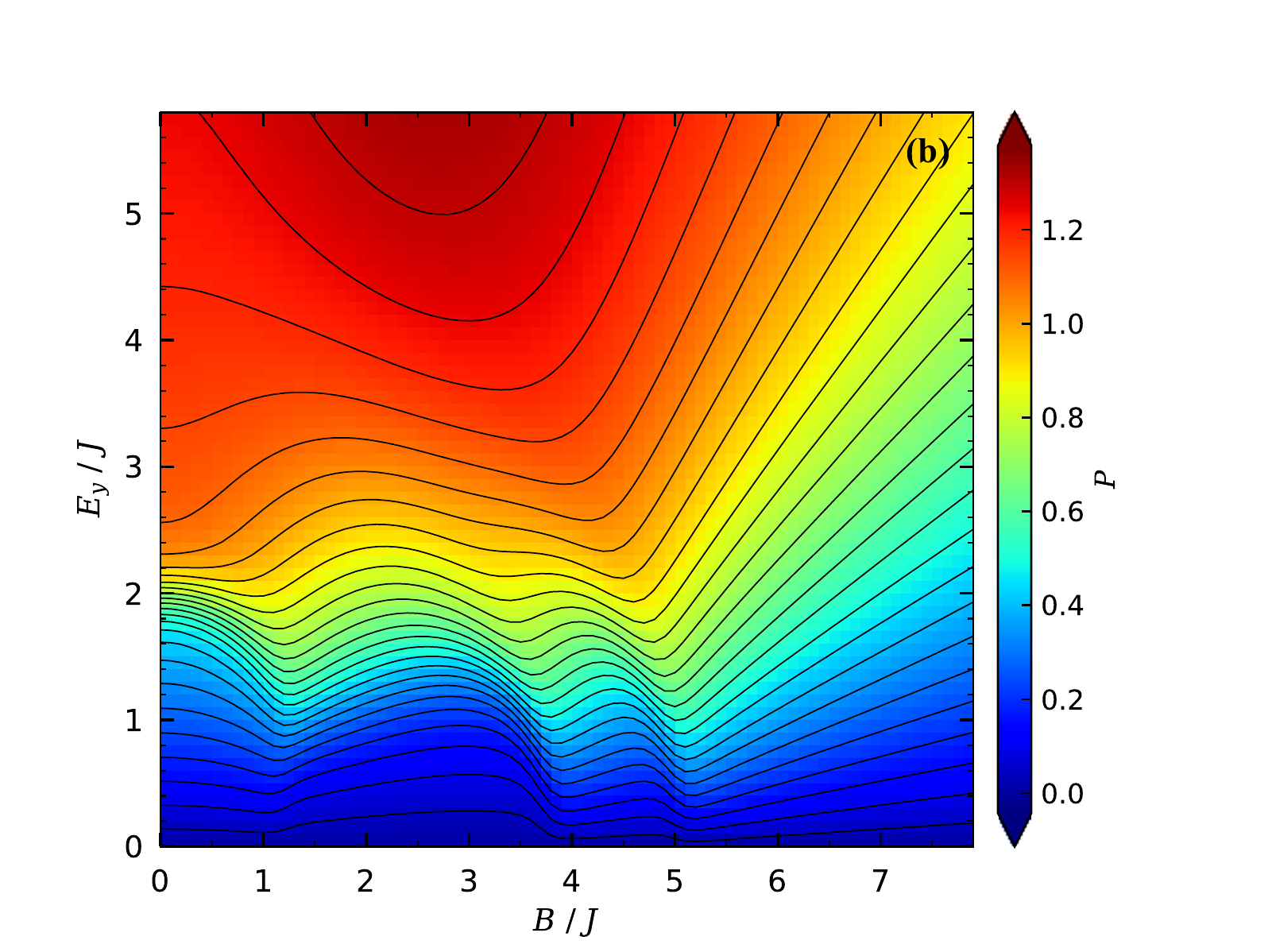}
}
\caption{ (a) Magnetization with respect to the saturation value of the Heisenberg XXZ octahedron model in the ($B/J-E_y/J$) plane at low temperature $T=0.06J_1$ and fixed $\Delta=2J$. (b) Polarization of the same model in ($B/J-E_y/J$) plane at low temperature $T=0.06J$ and fixed $\Delta=2J$. }
\label{fig:Mag_Pol_BEy_Oct}
\end{center}
\end{figure}

For a comparison, the magnetization process and polarization of the spin-1/2 XXZ Heisenberg octahedron (figure \ref{fig:SpinModel_T_O}(b)) at the finite low temperature $T=0.06J$ and fixed $\Delta=2J$ is shown in figure \ref{fig:Mag_Pol_BEy_Oct}. In an excellent agreement with the results of previous work \cite{str17a}, we see in  panel \ref{fig:Mag_Pol_BEy_Oct}(a) that the isothermal magnetization curve exhibits three well-known magnetization plateaus at zero, $(1/3)$ and $(2/3)$ of the saturation magnetization. It quite noteworthy that all magnetization plateaus gradually diminish upon increasing the electric field, and they are almost indiscernible in the magnetization curve in the presence of strong electric field. Also, the electric field increment causes the magnetization of the model reaches its saturation value in sufficiently strong magnetic fields.

The polarization data presented in panel \ref{fig:Mag_Pol_BEy_Oct}(b) would suggest that the polarization has sharp peaks close to the critical magnetic fields $B_c=J$, $B_c=3.75J$ and $B_c=5J$, at which magnetization jumps occur between the corresponding plateaus. With increase of the electric field $E_y/J$ the polarization increases and its curve becomes smooth even at the mentioned critical magnetic fields. This scenario reveals the diminishing magnetization plateaus upon increasing the electric field. The magnetic field changes result in changing the critical electric fields position at which giant ramp occurs in the polarization curve (see yellow region of panel  \ref{fig:Mag_Pol_BEy_Oct}(b)).

\begin{figure}
\begin{center}
\resizebox{0.45\textwidth}{!}{%
\includegraphics{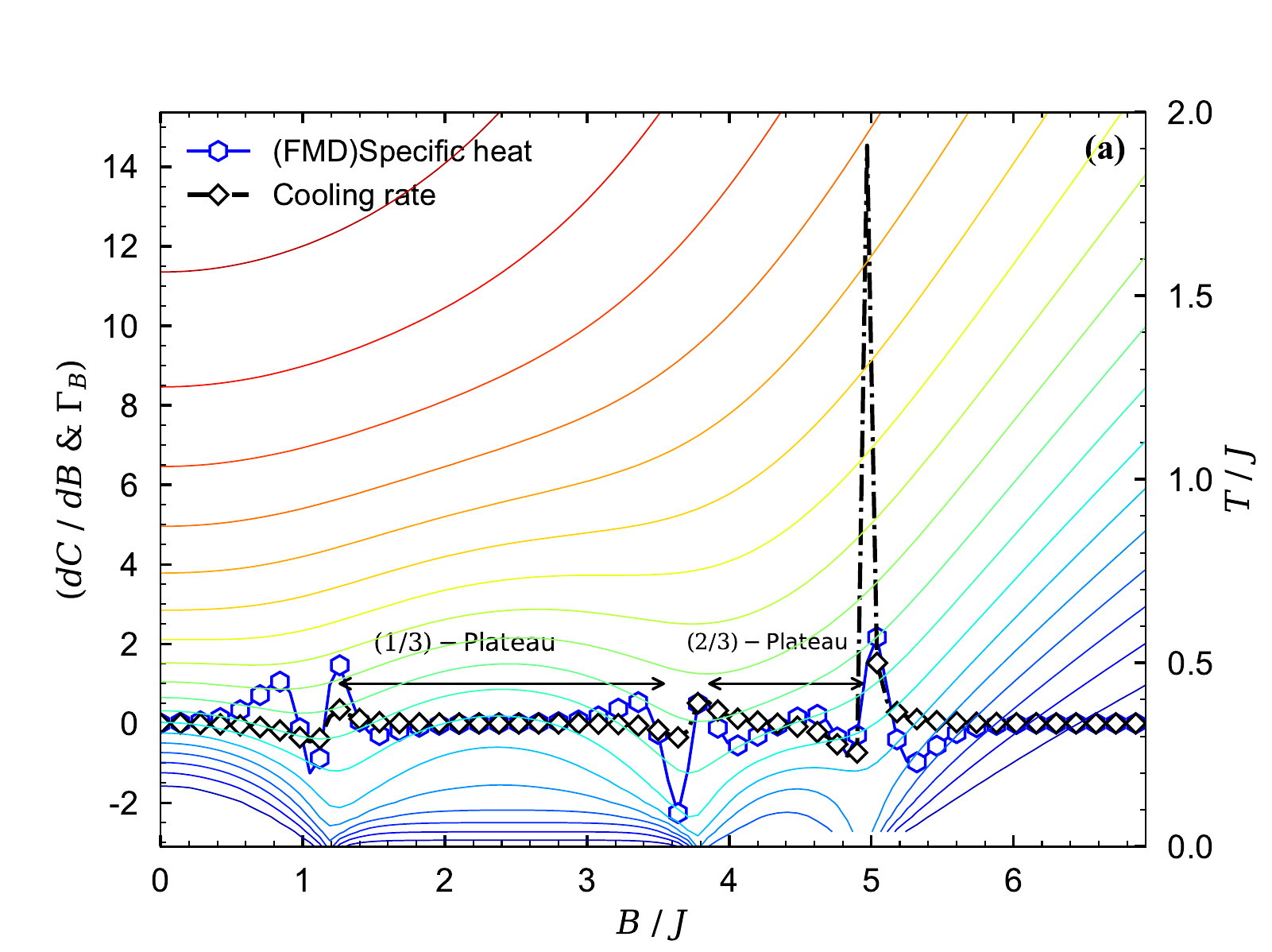}
}
\resizebox{0.45\textwidth}{!}{%
\includegraphics{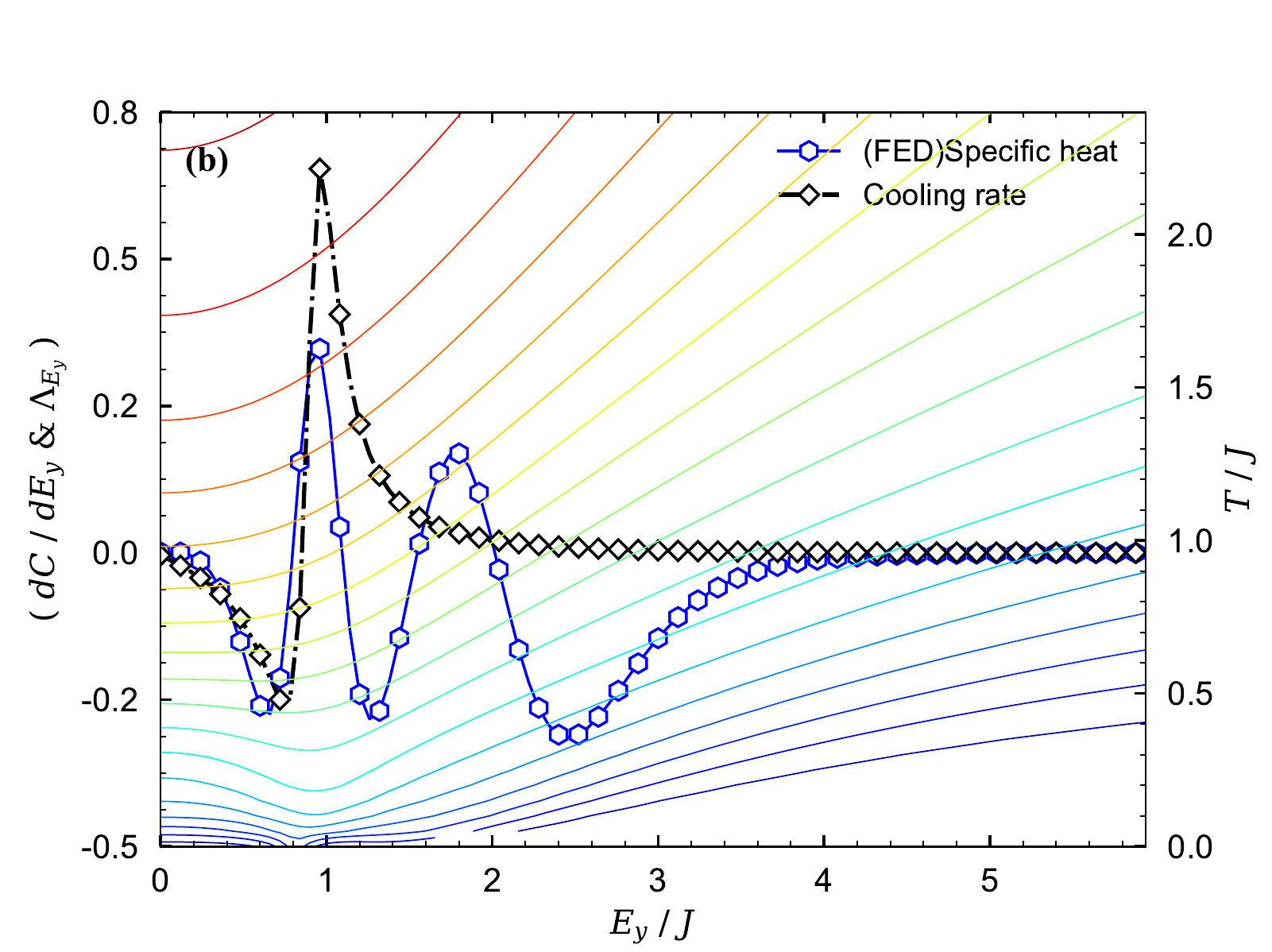}
}
\caption{(a) The first magnetic field derivative (FMD) of the specific heat and the cooling rate of the Heisenberg XXZ octahedron model  as functions of the magnetic field at sufficiently low temperature $T=0.06J$ for a set of other parameters as  $E_y=0.5J$, $\Delta=J_2=2J$. (b) The first electric field derivative (FED) of the specific heat and the cooling rate of the same model as functions of the electric field at sufficiently low temperature $T=0.06J$ for fixed values of $B=8J$, $\Delta=J_2=2J$. Isentropy lines are also illustrated in the corresponding panels under the same conditions as other parameters.}
\label{fig:SHeatDiffGammaS_BEy_Oct}
\end{center}
\end{figure}

For completeness, we plot in figure \ref{fig:SHeatDiffGammaS_BEy_Oct}(a) the first magnetic field derivative of the specific heat $\frac{dC}{dB}$ and the cooling rate of the spin-1/2 XXZ Heisenberg octahedron at low temperature $T=0.06J$ for the fixed values of $E_y=0.5J$ and $\Delta=2J$.
Furthermore, the contour plot for the entropy as a function of the magnetic field for the limiting case $E_y=0.5J$ and $\Delta=2J$  is shown in this figure. The application of the electric field provides an additional possibility to tune MCE of the considered model, as well as, paves the way to study the ECE of the model in several circumstances. 
For the case of weak electric field, we observe three critical magnetic fields at which enhanced magnetocaloric effect happens. The application of
the electric field leads to enhanced magnetocaloric effect for the spin-1/2 XXZ Heisenberg octahedron when the anisotropy parameter changes. Therefore, the application of the electric field creates a possibility to govern the strength of the magnetoelectric effect in this case. The effect of the electric field on the entropy is also interesting to follow. From figure \ref{fig:SHeatDiffGammaS_BEy_Oct}(a), we realize that even for a small electric field the lines of constant entropy shows a strong dependence of the electric field variations. The application of the magnetic field changes the behavior of the electric field dependence of the cooling rate from heating to cooling.
In the contrary of spin-1/2 XXZ Heisenberg two edge-shared tetrahedra model, the existence of zero magnetization plateau in a magnetization process of the spin-1/2 Heisenberg octahedron (panel \ref{fig:SHeatDiffGammaS_BEy_Oct}(a)), independent of the aniostropy and electric field strength, causes the absence of magnetocaloric effect in a proximity of zero magnetic field.

The adiabatic depolarization of the spin-1/2 octahedron model is characterized by a rapid decrease in temperature as the electric field diminishes  (panel \ref{fig:SHeatDiffGammaS_BEy_Oct}(b)). It should be stressed that the isentropy lines related to the low amounts of the entropy  reach almost the absolute zero temperature as the electric field tends to critical points $E_y=J$ and $E_y=2J$, which suggest the fastest possible cooling process that would be quite superior for this model. 
The variations of the function  $\frac{dC}{dE_y}$ clearly demonstrates existence of these two critical electric fields, at which the enhanced ECE  occurs. Therefore, a refrigeration with the help of adiabatic depolarization of the spin-1/2 XXZ octahedron model might be astonishingly useful in condensed matter for future technologies.
\section{Conclusion}\label{conclusion}
The present paper deals with magnetic, thermodynamic, magnetocaloric and electrocaloric properties of the two species of small spin clusters so-called: (I) two spin-1/2 XXZ Heisenberg edge-shared tetrahedra model, and (II)  spin-1/2 XXZ Heisenberg octahedron. We have assumed for the aforedescribed models a Dzyaloshinskii-Moriya type interaction between their interdimers, which denotes existence of a typical electric field.
All relevant quantities can be calculated rigorously and have been examined in detail using full exact numerical diagonalization method. Our exact results are in a good agreement with the previous works in the same conditions (\cite{kar17,str17a}).
It has been shown that the magnetization plateaus(polarization ramps) of the  two spin-1/2 XXZ Heisenberg edge-shared tetrahedra model rigorously depend on the external electric(magnetic) field. We also realized that the polarization in the $(\Delta/J_1-E_y/J_1)$ plane displays an intriguing ramp that its shape and position significantly depend on the strength of the magnetic field. Besides, we have noted that the specific heat of the tow spin-1/2 XXZ Heisenberg  edge-shared tetrahedra model has a Schottky-type maximum in the low magnetic fields that converts to a double-peak upon increasing the magnetic field. Interestingly, the shape of the double-peak quantitatively change by altering the electric field. These changes in the shape and position of the specific heat with respect to the fields variations are in a good coincidence with the magnetization jumps and polarization ramps.

The anisotropy influences on the magnetization and polarization of the two spin-1/2 XXZ Heisenberg edge-shared tetrahedra model have been investigated as well. Ultimately, we have found that the anisotropy property and the electric field has contrary effects on the magnetization and polarization behaviors. Furthermore, we have realized that the magnetization jumps accompanying a transition between different magnetization plateaus manifest themselves in an enhanced MCE accompanied by a relatively fast cooling of
system during the adiabatic (de)magnetization.  An enhanced ECE has been observed nearby a critical electric field that quantitatively depends on both of the magnetic field and anisotropy parameter.
From this point of view,  the spin-1/2 XXZ Heisenberg two edge-shared tetrahedra model has turned out as one of the most prominent small spin clusters that would allow an efficient cooling down at sufficiently low  temperatures.

For comparison,  the effect of electric field on the magnetization, as well as, the effect of magnetic field on the polarization of the spin-1/2 XXZ Heisenberg octahedron have been examined in detail. It has been demonstrated that the magnetization behavior with respect to the magnetic field is strongly dependent on the electric field as we found for the case spin-1/2 two edge-shared tetrahedra. The polarization of the spin-1/2 XXZ Heisenberg octahedron displayed anomalous behavior close to the critical magnetic fields, at which  field-driven quantum phase transitions accompanied with the magnetization jump occur. By altering the electric field, the polarization showed a giant ramp in a particular electric field. Therefore, we concluded that the polarization of the model can depict all magnetic phase boundaries in the $(E_y/J-B/J)$ plane. 

Our exact results of examining the isentropy data of the spin-1/2 Heisenberg octahedron obtained from the exact numerical method, detected in a low electric field region a steep decline of temperature upon diminishing of the magnetic field. From this perspective, theoretical realizations of the spin-1/2 Heisenberg octahedron can be also considered for an efficient low-temperature refrigeration, since the temperature decreased rapidly near critical magnetic(electric) fields due to enhanced magnetocaloric(electrocaloric) effects.
For the both models we discovered that the cooling rate coincide well with the first magnetic(electric) field derivative of the specific heat. However, we showed that the first electric field derivative of the specific heat is more sensitive nearby the critical electric fields at which enhanced magnetocaloric effect is observed.

Finally, the considered spin models hold promise as core systems permitting to examine some other aspects of magnetic and thermodynamic properties, e.g., the dynamical magnetization and dynamical magnetocaloric (electrocaloric) effect etc., that would be the subject of our future works.
\section*{Acknowledgments}
H. Arian Zad and N. Ananikian  acknowledge the receipt of the grant from the ICTP Affiliated Center Program AF-04 and the CS MES RA in the frame of the research project No. SCS 18T-1C155. We are also grateful to Vadim Ohanyan for insightful discussions.


\section*{References}\label{references}
\begin{thebibliography}
\bibliography{}

\bibitem{str18b} 
Stre{\v c}ka J and Kar{\v l}ova K 2018 {\it AIP Advances} \href{https://dx.doi.org/10.1063/1.5042029} {{\bf 8}, 101403}

\bibitem{str17a} 
 Kar{\v l}ova K and Stre{\v c}ka J and Richter J 2017 {\it J. Phys.: Condens. Matt.} \href{https://dx.doi.org/10.1088/1361-648X/aa53ab}  { {\bf 29}, 125802}

\bibitem{Martinez2013} 
Martinez G, Tangarife E, Perez M and Mejia-Lopez J 2013 {\it J. Phys.: Conde. Matt.} \href{https://doi:10.1088/0953-8984/25/21/216003}{{\bf 25}, 216003}

\bibitem{Die2016} 
Die D, Zheng BX, Zhao LQ, Zhu QW and Zhao ZQ 2016 {\it Phys. Rev. Lett.} \href{https:// DOI: 10.1038/srep31978}{{\bf 6}, 31978}     

 \bibitem{str15} 
 Stre{\v c}ka J   Kar{\v l}ova T and  Madaras T 2015 {\it Phys. B} \href{10.1016/j.physb.2015.03.031} {{\bf 466}, 76}

\bibitem{kar17} 
Kar{\v l}ova K and Stre{\v c}ka J 2017 {\it J. Low Temp. Phys.} \href{https://dx.doi.org/10.1007/s10909-016-1676-8} {{\bf 187}, 727}



\bibitem{Dmitriev12002} 
Dmitrieva  D V, Krivnova V Ya, Ovchinnikova A A, and Langari A 2002 {\it JETP} \href{https://doi.org/10.1134/1.1513828} {{\bf 95}, 538-549}
 
\bibitem{Dmitriev2004} 
Dmitriev D V, Krivnov V Ya 2004 {\it Phys. Rev. B} \href{https://doi.org/10.1103/PhysRevB.70.144414 }{{\bf 70}, 14, 144414}

\bibitem{Hieida2001} 
Hieida Y, Okunishi K, Akutsu Y 2001 {\it Phys. Rev. B} \href{https://https://doi.org/10.1103/PhysRevB.64.224422}{{\bf 64}, 22, 224422} 

\bibitem{Mori1995} 
Mori S, Kim J J, Harada I 1995 {\it J. Phys. Soc. Jap.} \href{ https://doi.org/10.1143/jpsj.64.3409}{{\bf 64} , 9, 3409-3415}  

 \bibitem{Dmitriev22002} 
Dmitriev D V, Krivnov V Ya, Ovchinnikov A A 2002 {\it Phys. Rev. B} \href{https://doi.org/10.1103/PhysRevB.65.172409 }{{\bf 65}, 17, 172409}

\bibitem{Hagemans2005} 
Hagemans R, Caux J-S, Low U 2005 {\it Phys. Rev. B} \href{https://doi.org/10.1103/PhysRevB.71.014437 }{{\bf 71}, 1, 014437}

\bibitem{Hikihara1998} 
Hikihara T, Furusaki A 1998 {\it Phys. Rev. B} \href{https://doi.org/10.1103/PhysRevB.58.R583}{{\bf 58}, 2}

\bibitem{Hikihara2004} 
Hikihara T, Furusaki A 2004 {\it Phys. Rev. B} \href{https://doi.org/10.1103/PhysRevB.69.064427}{{\bf 69}, 6, 064427} 

 \bibitem{Kurmann1982} 
Kurmann J, Thomas  H, Muller G 1982 {\it Phys. A: Stat. Mech. and App.} \href{doi:10.1016/0378-4371(82)90217-5}{{\bf 112}, 1, 235–255}

\bibitem{Ovchinnikov2003} 
Ovchinnikov  A A, Dmitriev D V, Krivnov V Ya and Cheranovskii V O 2003 {\it Phys. Rev. B} \href{https://doi.org/10.1103/PhysRevB.68.214406 }{{\bf 68}, 21, 214406}

 \bibitem{Yang11966} 
 Yang C N, Yang C P 1966 {\it Phys. Rev.} \href{https://doi.org/10.1103/PhysRev.150.321}{{\bf 150}, 1, 321-327}

\bibitem{Yang21966} 
Yang  C N, Yang C P 1966 {\it Phys. Rev.} \href{https://doi.org/10.1103/PhysRev.150.327}{{\bf 150}, 1, 327-339}

\bibitem{Caux2003} 
Caux J-S, Essler  F H L, Low U 2003 {\it Phys. Rev. B} \href{https://doi.org/10.1103/PhysRevB.68.134431}{{\bf 68} ,13, 134431}

\bibitem{Honecker2004} 
Honecker A, Schulenburg J and Richter J 2004 {\it J. Phys.: Conde. Matt.} \href{https://doi.org/10.1088/0953-8984/16/11/025}{{\bf 16}, S749-S758}

\bibitem{Lacroix2011} 
Lacroix C, Mendels P, Mila F 2011 Introduction to Frustrated Magnetism, (Berlin Heidelberg Springer-Verlag)

\bibitem{Bellucci2014} 
Bellucci S, Ohanyan V, Rojas O 2014 {\it Eur. Phys. Lett.} \href{https://doi.org/10.1209/0295-5075/105/47012}{{\bf 105}  47012} 

\bibitem{Ohanyan2015} 
Ohanyan V, Rojas O, Strecka J, Bellucci S 2015 {\it Phys. Rev. B} \href{https://doi.org/10.1103/PhysRevB.92.214423}{{\bf 92} 214423}

\bibitem{Ohanyan2018} 
Torrico J, Ohanyan V and Rojas O 2018 {\it J. Magn. Magn. Mat.} \href{https://doi.org/10.1016/j.jmmm.2018.01.044}{{\bf 454}, 85}

 \bibitem{Schulenburg2002} 
Schulenburg J, Honecker A, Schnack J, Richter J and Schmidt H-J 2002 {\it Phys. Rev. Lett.} \href{https://doi.org/10.1103/PhysRevLett.88.167207}{{\bf 88}, 167207}

\bibitem{Shapira2002} 
Shapira Y, Bindilatti V 2002 {\it J. Appl. Phys.} \href{ https://doi.org/10.1063/1.1507808}{{\bf 92}, 4155}  

\bibitem{Antonosyan2009} 
Antonosyan D, Bellucci S, Ohanyan V 2009 {\it Phys. Rev. B} \href{https://doi.org/10.1103/PhysRevB.79.014432}{{\bf 79}, 014432}

\bibitem{Paulinelli2013} 
Paulinelli H G, Souza S M D and Rojas O 2013 {\it J. Phys. Conde. Matt.} \href{https://doi.org/10.1088/0953-8984/25/30/306003}{{\bf 25}, 306003}  

\bibitem{Lisnyi2016} 
Lisnyi B, Stre{\v c}ka J 2016 {\it Phys. A} \href{https://doi.org/10.1016/j.physa.2016.06.088}{{\bf 462}, 104}

\bibitem{Gu2007} 
Gu B, Su G 2007 {\it Phys. Rev. B} \href{https://doi.org/10.1103/PhysRevB.75.174437}{{\bf 75}, 174437}  

\bibitem{Rojas2012} 
Rojas O, Rojas  M, Ananikian N S, Souza S M D 2012 {\it Phys. Rev. A} \href{https://doi.org/10.1103/PhysRevA.86.042330}{{\bf 86}, 1, 042330} 

\bibitem{Torrico2014} 
Torrico J, Rojas M, Souza S M D, Rojas O and Ananikian N S 2014 {\it  Eur. Phys. Lett.} \href{https://doi.org/10.1209/0295-5075/108/50007}{{\bf 108}, 50007}  

\bibitem{Ananikian2012} 
Ananikian N S, Ananikyan L N, Chakhmakhchyan L A and Rojas O {\it J. Phys. Condens. Matt.} \href{https://doi.org/10.1088/0953-8984/24/25/256001}{{\bf 24}, 256001}

\bibitem{Abgaryan2015} 
Abgaryan V S, Ananikian N S, Ananikyan L N, Hovhannisyan V V 2015 {\it Solid State Comm.} \href{https://doi.org/10.1016/j.ssc.2015.10.003}{{\bf 224}, 15}

\bibitem{Rojas2017} 
Rojas O, Rojas M, Souza S M D, Torrico J, Strecka J and Lyra M L 2017 {\it Phys. A} \href{https://doi.org/10.1016/j.physa.2017.05.099}{{\bf 486}, 367} 

 \bibitem{Koga1988} 
Koga A, Kumada S, Kawakami N and Fukui T 1988 {\it J. Phys. Soc. J.} \href{https://doi.org/10.1143/JPSJ.67.622}{{\bf 67}, 622} 

\bibitem{str16} 
J. Stre{\v c}ka, R. C. Al{\v e}cio, M. L. Lyra, O. Rojas 2016 {\it J. Magn. Magn. Mat.} \href{https://doi.org/10.1016/j.jmmm.2016.02.095}{{\bf 409}, 124} 

 \bibitem{Zad2017} 
 H. Arian  Zad, and N.  Ananikian 2017 {\it J. Phys.: Condens. Matt.}  \href{https://doi.org/10.1088/1361-648X/aa8dd0}{{\bf 29}, 455402} 

\bibitem{Zad2018} 
 H. Arian  Zad, and N.  Ananikian 2018 {\it J. Phys.: Condens. Matt.} \href{https://doi.org/10.1088/1361-648X/aab644} {{\bf 30}, 165403}


\bibitem{Volkova2018} 
L. M. Volkova and D. V. Marinin 2018 {\it  J. Phys.: Condens. Matt.} \href{https://doi.org/10.1088/1361-648X/aade0b}{{\bf 30}, 425801} 

\bibitem{Volkova2017} 
L. M. Volkova and D. V. Marinin 2017{\it  J. Supercond. Nov. Magn.} \href{https://doi.org/10.1007/s10948-016-3892-5}{{\bf 30}, 959} 

\bibitem{Volkova12018} 
L. M. Volkova and D. V. Marinin 2018 {\it  Phys. Chem. Min.} \href{https://doi.org/10.1007/s00269-018-0950-5}{{\bf 45}, 655}

\bibitem{Tishin16} 
Tishin  A M and Spichkin Y  I, (CRC Press, 2016).

\bibitem{Franco16} 
 Franco V, Bl{\' a}zquez J S, Ingale B and Conde A  2012 {\it  Ann. Rev. Mat. Re.} {{\bf 42} 305}.
 
 \bibitem{Correia14} 
Correia T and  Zhang Q, {\it Electrocaloric Effect: An Introduction.} \href{https://doi.org/10.1007/978-3-642-40264-7_1}{(Springer Berlin Heidelberg, 2014).}

\bibitem{Szalowski18} 
 Szalowski K and Balcerzak Y, \href{https://doi.org/10.1038/s41598-018-23443-x }{(2018) {\it Sci. Rep.} {\bf 8} 5116.}

 \bibitem{Littlewood16} 
Guzm{\' a}n-Verri G G and Littlewood P B 2016 {\it APL Mater.} \href{10.1063/1.4950788}{{\bf 4} 064106.}


\bibitem{Euler}
Euler L, 1768 {\it Institutionum calculi integralis}, (imp. Acad. imp. Sa{\' e}nt.)
 

\end{thebibliography}
\end{document}